\documentclass[review]{elsarticle}
\usepackage{setspace}
\usepackage{enumerate}
\usepackage{graphicx}
\usepackage{amsmath, bm}
\usepackage{float}
\usepackage{mdframed}
\usepackage{multirow}
\usepackage{textcomp}
\usepackage{subcaption}
\usepackage{tikz}
\usepackage{float}
\usetikzlibrary{
arrows.meta,
positioning,
shapes.geometric,
fit,
calc
}

\usepackage{amsmath}
\usepackage{url}
\usepackage{makecell}
\usepackage{amssymb}
\usepackage{bm}
\usepackage{textcmds}
 \usepackage{hyperref}
\biboptions{numbers,sort&compress}
\usepackage{tikz}
\usetikzlibrary{arrows.meta, positioning, shapes.geometric}
\usetikzlibrary{calc}
\tikzstyle{nnblock} = [
    draw, dashed, rounded corners,
    minimum width=5cm, minimum height=4cm
]

\tikzstyle{physblock} = [
    draw, dashed, rounded corners,
    minimum width=5cm, minimum height=4cm
]

\tikzstyle{neuron} = [
    draw, circle, fill=orange!25,
    minimum size=8mm
]

\tikzstyle{param} = [
    draw, circle, fill=purple!25,
    minimum size=8mm
]

\tikzstyle{loss} = [
    draw, ellipse, fill=green!20,
    minimum width=2.8cm
]

\usepackage{pgfplots}
\pgfplotsset{compat=1.18}
\usepackage{array}

\usetikzlibrary{shapes.geometric, arrows}
\usepackage{caption}  

\def\be{\begin{equation}}\def\ee{\end{equation}}
\def\ba{\begin{array}}\def\ea{\end{array}}
\def\bfg{\begin{figure}}\def\efg{\end{figure}}
\makeatletter
\def\fps@figure{htbp}
\makeatother

\usepackage{stackengine}
\stackMath
\newcommand\tenq[2][1]{%
 \def\useanchorwidth{T}%
  \ifnum#1>1%
    \stackunder[0pt]{\tenq[\numexpr#1-1\relax]{#2}}{\scriptscriptstyle\sim}%
  \else%
    \stackunder[1pt]{#2}{\scriptscriptstyle\sim}%
  \fi%
}

\RequirePackage[margin=20mm]{geometry}

\journal{}
\begin{document}

\begin{frontmatter}

\title{Fractional Modeling of Thermoelastic Fracture Behavior in a Cracked PZT-4 Strip under Transient Thermal Loading}
\author[label1]{Diksha}
\author[label1]{Soniya Chaudhary*}
\author[label1]{Pawan Kumar Sharma}
\cortext[cor1]{Corresponding author: soniyachaudhary18@gmail.com}
\address[label1]{Department of Mathematics and Scientific Computing, National Institute of Technology Hamirpur, Himachal Pradesh, 177005, India}

\begin{abstract}
This paper investigates the thermoelastic fracture response of a transversely isotropic piezoelectric strip containing a vertical insulated crack under transient thermal shock loading and pre-existing stress fields. The analysis is conducted within the framework of generalized fractional heat conduction using the Ezzat model, which incorporates thermal relaxation and memory-dependent effects.
The problem is formulated as a mixed boundary value problem governed by fractional thermoelastic equations. The Laplace transform technique is employed to obtain temperature and coupled fields in the transform domain. The resulting system of singular integral equations is solved using the Lobatto-Chebyshev collocation method to determine the displacement discontinuity and the associated thermal stress intensity factors at the crack tips. The transient response in the time domain is recovered through numerical inversion of the Laplace transform using the Stehfest algorithm.
Numerical results for PZT-4 are presented to examine the influence of fractional order, thermal relaxation time, pre-existing stresses, and geometric parameters on temperature distribution, thermoelastic stress fields, and stress intensity factors. The results demonstrate significant deviations from classical Fourier predictions, revealing wave-like thermal behavior and inherent memory effects associated with fractional heat conduction.
The present formulation establishes a unified framework for the analysis of 
thermoelastic fracture in piezoelectric ceramics and provides insights into 
the design and reliability of smart structures operating under severe thermal conditions.
\end{abstract}


\end{frontmatter}
\section{Introduction}
Electrical power generation in modern aerospace systems is traditionally achieved through mechanical extraction from turbine-driven sources. These approaches, while widely adopted, introduce efficiency losses due to mechanical transmission and impose additional load on the propulsion system. Furthermore, reliance on conventional energy storage devices, such as batteries, limits long-term operation and increases maintenance requirements, particularly in inaccessible or harsh environments.
To address these limitations, energy harvesting techniques have been proposed as sustainable alternatives for powering distributed and embedded systems. Among these, piezoelectric materials are of particular interest due to their ability to directly convert mechanical vibrations into electrical energy, making them well-suited for aerospace structures subjected to dynamic loading conditions \cite{le2015review,bowen2014pyroelectric}.
Piezoelectric materials exhibit coupled electromechanical behavior due to their non-centrosymmetric crystal structure, giving rise to the direct and converse piezoelectric effects. These properties form the basis for sensing, actuation, and energy harvesting applications. In particular, piezoelectric ceramics such as lead zirconate titanate (PZT) provide high electromechanical coupling and are widely used in aerospace systems \cite{elahi2020review}.

Despite these advantages, piezoelectric materials are inherently brittle and susceptible to defects such as microcracks introduced during fabrication or service. In aerospace environments, thermal loads arise from aerodynamic heating, engine proximity, and cyclic temperature variations. Under thermo-electro-mechanical loading, these effects generate thermal strains due to temperature gradients and material mismatch, leading to localized stress concentrations near defects. The interaction of thermal, mechanical, and electrical fields further amplifies these stresses, accelerating crack initiation and propagation and compromising structural integrity.
Consequently, the analysis of thermoelastic fracture in piezoelectric materials is essential for predicting failure and ensuring reliability in aerospace applications. Early studies include Rizk and Radwan \cite{rizk1993fracture}, who investigated fracture under transient thermal stresses, and Zhi-He and Noda \cite{zhi1994transient}, who derived stress intensity factors for cracks in functionally graded materials. Erdogan and Wu \cite{erdogan1996crack} further examined crack behavior in FGM layers, emphasizing the role of material gradation.
With the inclusion of electromechanical coupling, Wang and Mai \cite{wang2002cracked} analyzed transient thermal fracture in piezoelectric materials. Ueda \cite{ueda2007thermal} introduced thermal intensity factors for cracks in functionally graded piezoelectric strips, later extending the analysis to transient thermal loading \cite{ueda2008cracked}.
Additional studies have further examined thermo-electro-mechanical behavior and fracture characteristics in piezoelectric and functionally graded materials. Xiang and Shi \cite{xiang2009static} analyzed functionally graded piezoelectric structures under combined electro-thermal loading, providing insight into coupled field responses. Zhang and Wang \cite{zhang2012contact} investigated crack behavior in piezoelectric materials under thermo-mechanical loading using a contact zone approach, highlighting the influence of electrical boundary conditions on fracture response. 
Subsequently, Sladek et al. \cite{sladek2014fracture} studied fracture behavior in piezoelectric semiconductors under thermal loading, extending analysis to materials with coupled electrical and semiconductor effects. Ma et al. \cite{ma2016thermo} examined thermo-mechanical coupling in piezoelectric actuator-driven systems, demonstrating the role of thermal effects in device performance and reliability. Jiang et al. \cite{jiang2018numerical} conducted numerical analyses of residual stresses and crack propagation in thermal barrier coatings under cyclic thermal loading, emphasizing the importance of thermal fatigue and damage evolution in layered materials. More recent studies have considered increasingly complex configurations, including laminated structures with interfacial cracks \cite{ueda2018transient}, defect interactions in piezoelectric coatings \cite{hu2021interaction}, and curved crack geometries under general thermal loading \cite{nourazar2023fracture}.
Most of these studies are formulated within the framework of classical thermoelasticity based on Fourier's law of heat conduction. Although mathematically convenient, this approach assumes an infinite speed of heat propagation, which may not be realistic under high-rate thermal loading or at small scales.

To address the limitations of classical heat conduction, which assumes instantaneous thermal propagation, Cattaneo and Vernotte \cite{cattaneo1958form} introduced a hyperbolic heat conduction model incorporating a finite thermal wave speed. Building on this framework, Tzou \cite{tzou1995unified} proposed the dual-phase-lag (DPL) model, introducing distinct relaxation times for the heat flux and temperature gradient. These models provide improved descriptions of transient thermal processes, particularly in microscale and high-rate loading conditions where classical assumptions break down.
Further generalizations have been developed using fractional-order formulations to capture hereditary and memory-dependent effects in thermopiezoelectric and viscoelastic materials. Youssef \cite{youssef2010theory} and Sherief et al. \cite{sherief2010fractional} introduced fractional heat conduction models based on Caputo derivatives, while Povstenko \cite{povstenko2011fractional,povstenko2020fractional} employed Riemann-Liouville operators to analyze thermal responses in cracked media. Ezzat \cite{ezzat2011magneto} applied fractional formulations to wave propagation problems, and subsequent studies by Carpinteri et al. \cite{carpinteri2011fractional} and Rangelov et al. \cite{rangelov2018dynamic} demonstrated the effectiveness of fractional calculus in modeling crack-tip behavior and stress wave attenuation in graded media.
These theoretical developments have directly influenced fracture analysis in coupled systems. Wang and Li \cite{wang2013hyperbolic} investigated transient thermal fracture in a piezoelectric layer using hyperbolic heat conduction, demonstrating the importance of relaxation effects. Chen and Hu \cite{chen2014thermoelastic} further analyzed thermoelastic behavior in cracked layered structures within this framework.
More recent studies have applied advanced non-Fourier models to complex configurations. Zhou et al. \cite{zhou2019transient} examined transient thermo-electro-elastic contact problems in functionally graded piezoelectric materials, while Zhang et al. \cite{zhang2019transient} and Mondal et al. \cite{mondal2019transient} incorporated fractional and memory-dependent formulations to capture transient responses in cracked media.
Further developments include the work of Yang et al. \cite{yang2022dynamic}, who analyzed dynamic fracture in thermopiezoelectric materials under fractional heat conduction, and Yang et al. \cite{yang2024transient}, who employed the Guyer-Krumhansl model to capture size-dependent and nonlocal thermal effects in fracture problems.
In addition to these analytical developments, computational approaches such as meshfree and finite element methods have been employed to study fracture behavior in graded and piezoelectric materials \cite{meng2015enriched,li2015sbfem}. These studies collectively highlight the importance of incorporating non-Fourier heat conduction for realistic modeling of thermoelastic fracture processes.

 Despite the extensive body of work on thermoelastic fracture in piezoelectric and functionally graded materials, several important limitations remain. Early studies on thermal fracture \cite{rizk1993fracture,zhi1994transient,erdogan1996crack} and subsequent investigations incorporating electromechanical coupling \cite{wang2002cracked,ueda2007thermal,ueda2008cracked,ueda2018transient} have primarily been conducted within the framework of classical Fourier heat conduction. Although these models provide useful insights, they rely on the assumption of instantaneous thermal propagation, which limits their applicability under rapid thermal loading conditions such as thermal shock.
Recent developments in non-Fourier heat conduction, including hyperbolic, fractional, and nonlocal models \cite{wang2013hyperbolic,chen2014thermoelastic,zhou2019transient,zhang2019transient,mondal2019transient,yang2022dynamic,yang2024transient}, have improved the physical realism of thermal transport modeling. However, these studies are often restricted to simplified geometries or focus on specific aspects such as dynamic response or numerical implementation, without fully addressing fracture behavior under coupled thermo-electro-mechanical fields.
Moreover, existing investigations on fracture in piezoelectric and graded materials \cite{xiang2009static,zhang2012contact,sladek2014fracture,ma2016thermo,jiang2018numerical} generally neglect the combined effects of vertical crack configurations, transient thermal shock loading, and pre-existing mechanical stresses. In particular, the interaction between thermal shock-induced stress waves and initial stress fields in the presence of an insulated crack has not been adequately explored within a non-Fourier framework.
Therefore, a comprehensive formulation that simultaneously accounts for non-classical heat conduction, transient thermal shock, pre-stress conditions, electromechanical coupling, and realistic crack configurations in piezoelectric materials remains lacking. Addressing this gap is essential for accurately predicting fracture behavior and ensuring the reliability of piezoelectric structures operating under severe thermal environments.

To address the aforementioned limitations, the present study investigates the thermoelastic fracture behavior of a transversely isotropic piezoelectric strip containing a vertical crack under the combined effects of pre-existing stresses and transient thermal shock loading. The primary objective is to develop a comprehensive framework that captures the interaction between thermal shock, initial stress fields, and crack-induced stress concentrations within a non-classical heat conduction setting.
The analysis is carried out within the framework of fractional-order heat conduction, specifically employing the Ezzat model, which incorporates memory effects and thermal relaxation characteristics to provide a more physically realistic description of heat transport in coupled thermo-electro-mechanical systems. This approach enables accurate modeling of finite-speed thermal propagation and history-dependent behavior, which are essential under high-rate thermal loading conditions.
An analytical formulation is developed to derive the governing singular integral equations of the problem, which are solved using the Lobatto-Chebyshev collocation method to determine the displacement discontinuity at the crack tip. The transient response is obtained through numerical Laplace inversion based on the Stehfest algorithm. Closed-form and numerical results are obtained for the temperature distribution, thermal stress fields, and thermal stress intensity factors associated with the crack.
For numerical investigations, the piezoelectric material PZT-4 is considered due to its extensive use in aerospace structures, sensors, and smart systems, where coupled thermo-electro-mechanical effects are significant.
The novelty of this work lies in the unified treatment of fractional heat conduction, transient thermal shock, pre-existing stresses, and vertical crack configuration within a single framework. The results provide new insights into crack-tip behavior under non-classical thermal conditions and contribute to the design and reliability assessment of piezoelectric structures operating in severe thermal environments. The remainder of this manuscript is organized as follows. Section \ref{Mathematical Formulation} presents the mathematical formulation of the problem and the derivation of the thermal stress intensity factors. In particular, Subsection \ref{Problem Geometry} describes the geometry of the problem, while Subsection \ref{Fractional Heat Conduction Model} introduces the fractional heat conduction model. The temperature field is obtained in Subsection \ref{Temperature Field Solution}, followed by the analysis of thermoelastic fields in the uncracked strip in Subsection \ref{Thermoelastic Field in the Uncracked Strip}. The electromechanical fields and the derivation of the thermal stress intensity factors are presented in Subsection \ref{Electromechanical Fields}.
Section \ref{Numerical Results and Discussion} provides numerical results and discussion, including graphical illustrations. Finally, Section \ref{Conclusion} summarizes the main findings of the study.

\section{ Mathematical Formulation}
\label{Mathematical Formulation}
\subsection{Problem Geometry}
\label{Problem Geometry}
\begin{figure}
    \centering
    \includegraphics[width=0.9\linewidth]{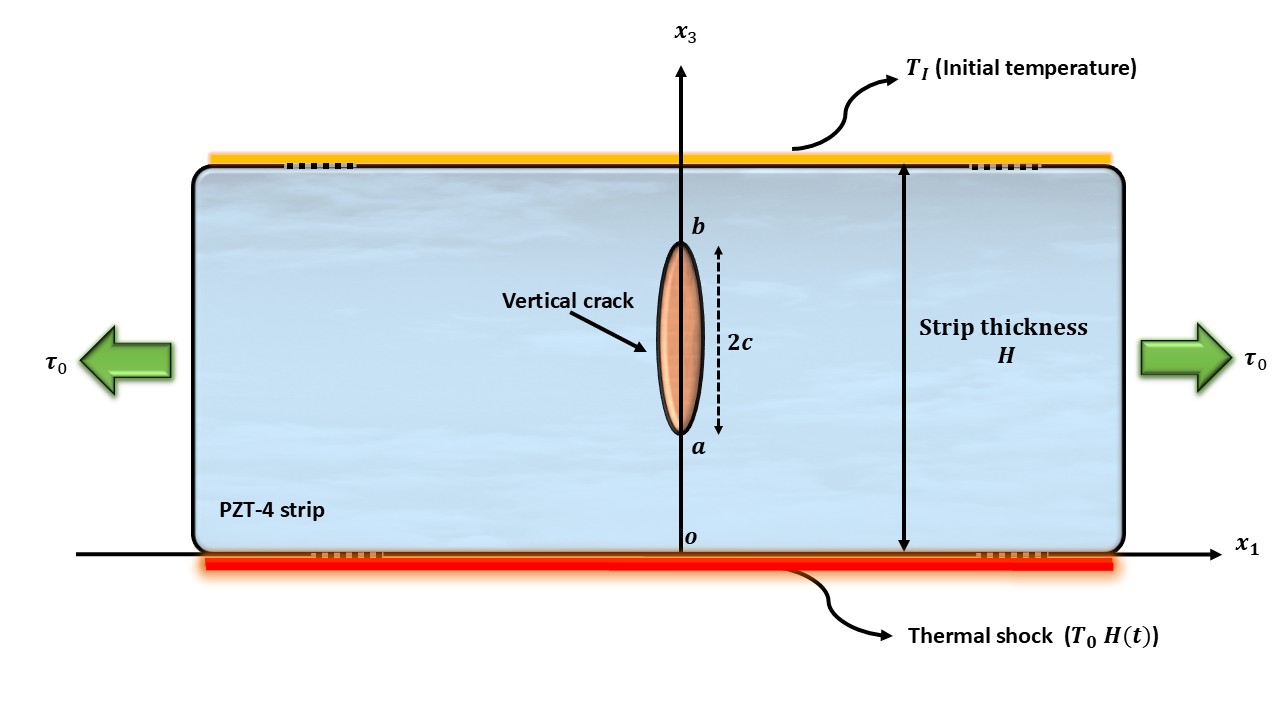}
   \caption{Schematic diagram of a PZT-4 piezoelectric strip containing an internal vertical crack of length $2c$}
    \label{GEOMETERY}
\end{figure}

A transversely isotropic piezoelectric strip of thickness $H$ is considered, 
containing an insulated crack of length $2c = b - a$, with $0 \le a < b < H$, 
as illustrated in Fig.~\ref{GEOMETERY}. The Cartesian coordinate system $(x_1,x_2,x_3)$ is introduced such that the $x_1$-axis lies along the lower boundary of the strip, while the $x_3$-axis is directed along the thickness of the strip. The crack is located along the $x_3$-axis within the region $a \le x_3 \le b$.
The piezoelectric medium is polarized along the $x_3$-direction and subjected to uniform initial stresses $\sigma_{11}^{0}$ and $\sigma_{33}^{0}$ acting along the $x_1$ and $x_3$ directions, respectively. Under these conditions, the problem is considered within the framework of plane strain in the $(x_1,x_3)$-plane, such that the displacement field is given by
\[
u_1 = u_1(x_1,x_3,t), \qquad u_2 = 0, \qquad u_3 = u_3(x_1,x_3,t),
\]
where $u_i$ $(i = 1,3)$ denote the displacement components along the 
corresponding coordinate directions.
The presence of the crack results in a Mode-I (opening mode) fracture 
configuration, characterized by the separation of crack faces under normal 
tensile stresses. Initially, strip is maintained at a uniform temperature 
$T_I$, taken as stress-free reference state. At time $t = 0$, the lower 
surface ($x_3 = 0$) is subjected to a sudden thermal loading of magnitude 
$T_0 H(t)$, where $H(t)$ denotes the Heaviside step function, while the 
upper surface ($x_3 = H$) remains at the initial temperature $T_I$. 
Throughout the transient process, the crack faces are assumed to be 
 electrically and thermally insulated.













\subsection{Fractional Heat Conduction Model}
\label{Fractional Heat Conduction Model}
A fractional-order heat conduction framework based on Ezzat’s formulation is employed to model heat transfer in piezoelectric media, incorporating finite-speed thermal propagation and memory effects.
In this formulation, the classical time derivative is replaced by a Caputo fractional derivative of order $\gamma$ $(0 < \gamma \leq 1)$, thereby extending the conventional Cattaneo--Vernotte model to capture memory-dependent effects. Accordingly, the governing non-Fourier constitutive relation for the heat flux vector $q_i$ is expressed as \cite{yu2020fractional}:
\begin{equation}
\left(1 + \frac{\tau_q^{\gamma}}{\Gamma(1+\gamma)}
\frac{\partial^{\gamma}}{\partial t^{\gamma}}\right) q_i
= -k_i\, T_{,i},
\label{eq:ezzat_heat_flux}
\end{equation}
where $\tau_q$ denotes the thermal relaxation time and $k_i$ represents the thermal conductivity coefficient along the $x_i$-direction.
The Caputo fractional derivative of order $\gamma$ with respect to time is defined as \cite{yu2020fractional}:
\begin{equation}
\frac{\partial^{\gamma} f(t)}{\partial t^{\gamma}}
=
\frac{1}{\Gamma(1-\gamma)}
\int_{0}^{t}
\frac{\partial f(\tau)/\partial \tau}
{(t-\tau)^{\gamma}}\, d\tau,
\qquad 0<\gamma<1.
\label{eq:caputo_definition}
\end{equation}
When $\gamma=1$, the fractional model reduces to the classical Cattaneo--Vernotte heat conduction model. The symbol $\Gamma(\cdot)$ denotes the Gamma function.
In the absence of internal heat generation, the energy balance equation governing the thermal behavior of the piezoelectric medium is given by \cite{yang2022dynamic}
\begin{equation}
\rho c_{\rho}\,\frac{\partial T}{\partial t}
= -\nabla \cdot \mathbf{q},
\label{eq:energy_equation}
\end{equation}
where $\rho$ and $c_{\rho}$ denote the mass density and the specific heat capacity of the material, respectively.

For the present problem, a one-dimensional transient heat conduction process is considered along the thickness direction. The temperature field is expressed as the sum of the initial uniform temperature $T_I$ and a perturbation component $T^{(1)}(x_3,t)$, i.e.,
\begin{equation}
T(x_3,t)=T_I+T^{(1)}(x_3,t),
\label{eq:temperature_decomposition}
\end{equation}
where $T_I$ denotes the initial (stress-free) temperature and $T^{(1)}(x_3,t)$ represents the transient temperature rise induced by the applied thermal loading.

Combining the energy balance equation without heat source~\eqref{eq:energy_equation} with the fractional non-Fourier heat conduction law~\eqref{eq:ezzat_heat_flux}, the non-uniform temperature field $T^{(1)}(x_3,t)$ satisfies the following fractional heat conduction equation:
\begin{equation}
\frac{\partial^2 T^{(1)}}{\partial x_3^2}
=
\frac{1}{\lambda_0}
\left(
\frac{\partial T^{(1)}}{\partial t}
+
\frac{\tau_q^{\gamma}}{\Gamma(1+\gamma)}
\frac{\partial^{\gamma+1} T^{(1)}}{\partial t^{\gamma+1}}
\right),
\label{eq:fractional_expanded_temperature}
\end{equation}
where $\lambda_0=\dfrac{k_3}{(\rho c_{\rho})}$ denotes the effective thermal diffusivity.
According to the considered problem, the initial and boundary conditions for the non-uniform temperature distribution 
$T^{(1)}(x_3,t)$ in the time domain are given by
\begin{align}
T^{(1)}(x_3,0) &= 0, 
\label{eq:IC} \\[4pt]
T^{(1)}(0,t) &= T_0\,H(t),
\label{eq:BC1} \\[4pt]
T^{(1)}(H,t) &= 0.
\label{eq:BC2}
\end{align}
\subsection{Temperature Field Solution}
\label{Temperature Field Solution}
The Laplace transform in time is employed to convert the fractional heat conduction equation into the Laplace domain, thereby simplifying the governing equations into an algebraic form. The corresponding transform pair used in the present analysis is defined as \cite{yang2022dynamic}:
\begin{equation}
f^{\ast}(x_1,x_3,s)
=
\int_{0}^{\infty}
f(x_1,x_3,t)\,e^{-st}\,dt,
\qquad
f(x_1,x_3,t)
=
\frac{1}{2\pi i}
\int_{\mathrm{Br}}
f^{\ast}(x_1,x_3,s)\,e^{st}\,ds,
\label{eq:Laplace_pair}
\end{equation}
where $s$ denotes the Laplace transform variable and the integration path $\mathrm{Br}$ represents the Bromwich contour in the complex $s$-plane.

The Laplace transform of the Caputo fractional derivative of order $\gamma$ $(0<\gamma\le1)$ is given by
\begin{equation}
\mathcal{L}\!\left\{
\frac{\partial^{\gamma} f(t)}{\partial t^{\gamma}}
\right\}
=
s^{\gamma} f^{\ast}(s)
-
\sum_{k=0}^{m-1}
f^{(k)}(0^{+})\, s^{\gamma-1-k},
\label{eq:Laplace_fractional}
\end{equation}
where $m$ is the largest integer less then  $\gamma$, and $f^{\ast}(s)$ denotes the Laplace transform of $f(t)$.

Applying the Laplace transform to the fractional heat conduction equation yields the governing differential equation for the transformed temperature increment $T^{(1)\ast}(x_3,s)$ as
\begin{equation}
\frac{1}{\lambda_0}
\left(
s+
\frac{\tau_q^{\gamma}}{\Gamma(1+\gamma)}s^{\gamma+1}
\right)
T^{(1)\ast}=\frac{d^{2}T^{(1)\ast}}{dx_3^{2}}.
\qquad 0<\gamma<1,
\label{eq:Laplace_temperature}
\end{equation}
The boundary conditions in the Laplace domain are
\begin{align}
T^{(1)\ast}(0,s) &= \frac{T_0}{s},
\qquad
T^{(1)\ast}(H,s) = 0.
\label{eq:Laplace_BC2}
\end{align}
Solving Eq.~\eqref{eq:Laplace_temperature} together with boundary conditions \eqref{eq:Laplace_BC2} yields  temperature increment in the Laplace domain as
\begin{equation}
T^{(1)\ast}(x_3,s)
=
\frac{T_0}{s}
\frac{
\exp(-\mu_0 x_3)
-
\exp[-\mu_0(2H-x_3)]
}{
1-\exp(-2\mu_0 H)
},
\label{eq:Laplace_temp_solution}
\end{equation}
where
\begin{equation}
\mu_0=
\left[
\frac{1}{\lambda_0}
\left(
s+
\frac{\tau_q^{\gamma}}{\Gamma(1+\gamma)}s^{\gamma+1}
\right)
\right]^{1/2}.
\label{eq:mu0}
\end{equation}
The transient temperature distribution is evaluated numerically by computing the inverse Laplace integral using the Stehfest numerical inversion algorithm.

\subsection{Thermoelastic Response of an Uncracked Piezoelectric Strip}
\label{Thermoelastic Field in the Uncracked Strip}
The temperature $T_I$ is adopted as reference temperature at which material is free of thermal stresses.
Consequently, the thermoelastic and pyroelectric fields considered in this section 
are generated solely by the non-uniform temperature increment $T^{(1)}(x_3,t)$. 
Once the temperature field is obtained from the fractional heat conduction equation, 
the associated thermoelastic stresses and electric displacements can be determined 
from the constitutive relations of a transversely isotropic piezoelectric medium \cite{ueda2008cracked}:
\begin{align}
\tau^{T}_{11} &= 
   \mu_{11}\frac{\partial u^{T}_{1}}{\partial x_1}
  +\mu_{13}\frac{\partial u^{T}_{3}}{\partial x_3}
  +e_{31}\frac{\partial \phi^{T}}{\partial x_3}
  -\kappa_{11}T^{(1)}, 
\label{eq:therm_tau11}
\\[6pt]
\tau^{T}_{33} &= 
   \mu_{13}\frac{\partial u^{T}_{1}}{\partial x_1}
  +\mu_{33}\frac{\partial u^{T}_{3}}{\partial x_3}
  +e_{33}\frac{\partial \phi^{T}}{\partial x_3}
  -\kappa_{33}T^{(1)}, 
\label{eq:therm_tau33}
\\[6pt]
\tau^{T}_{13} &= 
   \mu_{44}\!\left(
   \frac{\partial u^{T}_{1}}{\partial x_3}
   +
   \frac{\partial u^{T}_{3}}{\partial x_1}
   \right)
   +e_{15}\frac{\partial \phi^{T}}{\partial x_1},
\label{eq:therm_tau13}
\\[6pt]
D^{T}_{1} &= 
   e_{15}\!\left(
   \frac{\partial u^{T}_{1}}{\partial x_3}
   +
   \frac{\partial u^{T}_{3}}{\partial x_1}
   \right)
   -\varepsilon_{11}\frac{\partial \phi^{T}}{\partial x_1},
\label{eq:therm_D1}
\\[6pt]
D^{T}_{3} &= 
   e_{31}\frac{\partial u^{T}_{1}}{\partial x_1}
  +e_{33}\frac{\partial u^{T}_{3}}{\partial x_3}
  -\varepsilon_{33}\frac{\partial \phi^{T}}{\partial x_3}
  +p_z T^{(1)},
\label{eq:therm_D3}
\end{align}
where $\tau_{ij}^{T}$ denote the components of the thermally induced Cauchy stress tensor, and $D_i^{T}$ represent the components of the electric displacement vector (electric flux density) arising due to thermo-piezoelectric coupling. The quantities $u_1^{T}(x_3,t)$ and $u_3^{T}(x_3,t)$ denote the thermally induced displacement components, while $\phi^{T}(x_3,t)$ represents the thermally induced electric potential.  The superscript $T$ indicates that the corresponding quantities arise solely from the temperature increment $T^{(1)}$. 
The parameters $\mu_{ij}$ represent the elastic stiffness coefficients of the piezoelectric material, $e_{ij}$ denote the piezoelectric coupling constants, and $\varepsilon_{ij}$ are the dielectric permittivity coefficients. The constants $\kappa_{11}$ and $\kappa_{33}$ correspond to the thermoelastic stress coefficients, while $p_z$ denotes the pyroelectric constant that characterizes the coupling between temperature variation and electric displacement in the material.

Since the strip is considered infinite and its surfaces at $x_3=0$ and $x_3=H$ 
are free of traction and electrically insulated, the boundary conditions take the form:
\begin{equation}
\tau^{T}_{33}(x_3,t)=0,
\qquad
\tau^{T}_{13}(x_3,t)=0,
\qquad
D^{T}_{1}(x_3,t)=0,
\qquad
D^{T}_{3}(x_3,t)=0.
\label{eq:therm_BC}
\end{equation}
Under these conditions, all non-vanishing field quantities depend only on the thickness 
coordinate $x_3$. Solving Eqs.~\eqref{eq:therm_tau33}--\eqref{eq:therm_D3} for 
$\dfrac{\partial u^{T}_{3}}{\partial x_3}$ and 
$\dfrac{\partial \phi^{T}}{\partial x_3}$ 
and substituting the resulting expressions into Eq.~\eqref{eq:therm_tau11}, 
the axial thermoelastic stress can be expressed in the reduced form
\begin{equation}
\tau^{T}_{11}
=
\mu_{E0}\frac{\partial u^{T}_{1}}{\partial x_1}
-
k_{E0}T^{(1)}(x_3,t),
\label{eq:reduced_tau}
\end{equation}
where the effective elastic and thermal coefficients are defined as
\begin{equation}
\mu_{E0}
=
\mu_{11}^{0}
-
\frac{
\mu_{13}^{0}(\mu_{13}^{0}\varepsilon_{33}^{0}+e_{31}^{0}e_{33}^{0})
+
e_{31}^{0}(\mu_{13}^{0}e_{33}^{0}-\mu_{33}^{0}e_{31}^{0})
}{
\mu_{33}^{0}\varepsilon_{33}^{0}+(e_{33}^{0})^{2}
},
\label{eq:muE0}
\end{equation}
\begin{equation}
k_{E0}
=
\kappa_{11}^{0}
-
\frac{
\kappa_{33}^{0}(\mu_{13}^{0}\varepsilon_{33}^{0}+e_{31}^{0}e_{33}^{0})
-
p_z^{0}(\mu_{13}^{0}e_{33}^{0}-\mu_{33}^{0}e_{31}^{0})
}{
\mu_{33}^{0}\varepsilon_{33}^{0}+(e_{33}^{0})^{2}
}.
\label{eq:kE0}
\end{equation}
The compatibility condition for the present one-dimensional deformation reduces to
\begin{equation}
\frac{\partial^{2}}{\partial x_3^{2}}
\left(
\frac{\partial u^{T}_{1}}{\partial x_1}
\right)=0,
\label{eq:compat}
\end{equation}
which yields
\begin{equation}
\frac{\partial u^{T}_{1}}{\partial x_1}
=
A^{T}(t)x_3+B^{T}(t).
\label{eq:ux_general}
\end{equation}
Here $A^{T}(t)$ and $B^{T}(t)$ are time-dependent functions determined 
from the force-free boundary conditions of the strip.
Substituting Eq.~\eqref{eq:ux_general} into Eq.~\eqref{eq:reduced_tau} gives the 
axial thermoelastic stress distribution
\begin{equation}
\tau^{T}_{11}(x_3,t)
=
\mu_{E0}\left(A^{T}(t)x_3+B^{T}(t)\right)
-
k_{E0}T^{(1)}(x_3,t).
\label{eq:tau_final}
\end{equation}
Applying the Laplace transform with respect to time leads to
\begin{equation}
\tau^{T\ast}_{11}(x_3,s)
=
\mu_{E0}\left(A^{T\ast}(s)x_3+B^{T\ast}(s)\right)
-
k_{E0}T^{(1)\ast}(x_3,s).
\label{eq:tau_laplace}
\end{equation}
For an unconstrained strip, the resultant axial force and bending moment must vanish, 
leading to the conditions
\begin{equation}
\int_{0}^{H}\tau^{T\ast}_{11}(x_3,s)\,dx_3=0,
\qquad
\int_{0}^{H}x_3\,\tau^{T\ast}_{11}(x_3,s)\,dx_3=0.
\label{eq:force_moment}
\end{equation}
The stress distribution $\tau_{0}^{T*}(x_{3},s)=\tau^{T\ast}_{11}(x_3,s)$ obtained above 
is subsequently applied in the formulation of the cracked strip problem, serving 
as equivalent crack-surface traction.

\subsection{Electromechanical Fields}
\label{Electromechanical Fields}
The configuration illustrated in Fig. \ref{GEOMETERY} possesses a symmetry plane at \(x_{1}=0\) with respect to both the strip geometry and the applied thermo–electro–mechanical loading. Owing to this symmetry, the analysis can be restricted to the half-domain \(0 \leq x_{1} < \infty\), which significantly simplifies the mathematical formulation of the problem. 
By employing an appropriate superposition technique, the original boundary-value 
problem is transformed into an equivalent problem in which the crack-face 
tractions are the only non-zero external loads. Under this formulation, the resulting perturbation stresses must vanish as \(x_{1} \to \infty\), ensuring that the required far-field decay conditions are satisfied.
The piezoelectric strip is assumed to be subjected to uniform initial stresses \(\sigma_{11}^{0}\) and \(\sigma_{33}^{0}\) acting along the \(x_{1}\) and \(x_{3}\) directions, respectively. Under these conditions, the coupled electromechanical behavior of the piezoelectric medium is governed by the following field equations.
\begin{equation}
\mathbf{A}\,\frac{\partial^{2} \mathbf{u}}{\partial x_{1}^{2}}
+
\mathbf{B}\,\frac{\partial^{2} \mathbf{u}}{\partial x_{3}^{2}}
+
\mathbf{C}\,\frac{\partial^{2} \mathbf{u}}{\partial x_{1}\partial x_{3}}
=
\mathbf{0},
\label{eq:governing}
\end{equation}
where
\begin{equation}
\mathbf{u} =
\begin{bmatrix}
u_{1} \\
u_{3} \\
\phi
\end{bmatrix},
\end{equation}
and
\begin{equation}
\mathbf{A} =
\begin{bmatrix}
\mu_{11}+\sigma_{11}^{0} & 0 & 0 \\
0 & \mu_{44}+\sigma_{11}^{0} & e_{15} \\
0 & e_{15} & -\varepsilon_{11}
\end{bmatrix},
\,
\mathbf{B} =
\begin{bmatrix}
\mu_{44}+\sigma_{33}^{0} & 0 & 0 \\
0 & \mu_{33}+\sigma_{33}^{0} & e_{33} \\
0 & e_{33} & -\varepsilon_{33}
\end{bmatrix},
\,
\mathbf{C} =
\begin{bmatrix}
0 & \mu_{13}+\mu_{44} & e_{31}+e_{15} \\
\mu_{13}+\mu_{44} & 0 & 0 \\
e_{31}+e_{15} & 0 & 0
\end{bmatrix}.
\end{equation}
By applying the Laplace transform (\ref{eq:Laplace_pair}) with respect to time $t$ to Eqs.~(\ref{eq:governing}), the transformed field equations can be written in compact matrix form as
\begin{equation}
\mathbf{A}\,\frac{\partial^{2} \mathbf{u}^{*}}{\partial x_{1}^{2}}
+
\mathbf{B}\,\frac{\partial^{2} \mathbf{u}^{*}}{\partial x_{3}^{2}}
+
\mathbf{C}\,\frac{\partial^{2} \mathbf{u}^{*}}{\partial x_{1}\partial x_{3}}
=
\mathbf{0},
\label{eq:laplace_governing}
\end{equation}
where
\begin{equation}
\mathbf{u}^*(x_1,x_3,s) =
\begin{bmatrix}
u_{1}^*(x_1,x_3,s) \\
u_{3}^*(x_1,x_3,s) \\
\phi^*(x_1,x_3,s)
\end{bmatrix},
\end{equation}
with $u_1^*(x_1,x_3,s)$ and $u_3^*(x_1,x_3,s)$ denoting the Laplace transforms of the displacement components, and $\phi^*(x_1,x_3,s)$ denoting the Laplace transform of the electric potential.
The boundary conditions governing the cracked piezoelectric strip are given by:
\begin{align}
\tau_{11}(0,x_{3},t)
    &= -\tau_{0} - \tau_{0}^{T}(x_{3},t),
    \qquad  a < x_{3} < b,
\tag{BC1}\label{eq:BC_tau_xx}
\\[6pt]
u_{1}(0,x_{3},t)
    &= 0,
    \qquad 0 \le x_{3} \le a,\; b \le x_{3} \le H,
\tag{BC2}\label{eq:BC_ux_n}
\\[6pt]
\tau_{13}(0,x_{3},t)
    &= 0,
    \qquad 0 \le x_{3} \le H,
\tag{BC3}\label{eq:BC_tau_xz}
\\[6pt]
D_1(0,x_{3},t)
    &= 0,
    \qquad 0 \le x_{3} \le H.
\tag{BC4}\label{eq:BC_Dx}
\end{align}
and along the upper and lower surfaces of the strip,
\begin{equation}
\left.
\begin{aligned}
\tau_{13}(x_{1},0,t) &= 0, 
&\qquad 
\tau_{13}(x_{1},H,t) &= 0, \\[4pt]
\tau_{33}(x_{1},0,t) &= 0, 
&\qquad 
\tau_{33}(x_{1},H,t) &= 0, \\[4pt]
D_{3}(x_{1},0,t) &= 0,
&\qquad 
D_{3}(x_{1},H,t) &= 0,
\end{aligned}
\right\}
\qquad 0 \le x_{1} < \infty.
\tag{BC5}
\label{eq:BC_surfaces}
\end{equation}
Applying the Laplace transform defined in Eq.~(\ref{eq:Laplace_pair}) to the boundary conditions (\ref{eq:BC_tau_xx})-(\ref{eq:BC_surfaces}) yields the following transformed boundary conditions:
\begin{align}
\tau_{{1}{1}}^{*}(0,x_{3},s)
    &= -\frac{\tau_{0}}{s} - \tau_{0}^{T*}(x_{3},s),
    \qquad  a < x_{3} < b,
\tag{BC1$^*$}\label{eq:BC_tau_xx_star}
\\[6pt]
u_{1}^{*}(0,x_{3},s)
    &= 0,
    \qquad 0 \le x_{3} \le a,\; b \le x_{3} \le H,
\tag{BC2$^*$}\label{eq:BC_ux_star}
\\[6pt]
\tau_{{1}{3}}^{*}(0,x_{3},s)
    &= 0,
    \qquad 0 \le x_{3} \le H,
\tag{BC3$^*$}\label{eq:BC_tau_xz_star}
\\[6pt]
D_{{1}}^{*}(0,x_{3},s)
    &= 0,
    \qquad 0 \le x_{3} \le H,
\tag{BC4$^*$}\label{eq:BC_Dx_star}
\end{align}
and along the upper and lower surfaces of the strip,
\begin{equation}
\left.
\begin{aligned}
\tau_{{1}{3}}^{*}(x_{1},0,s) &= 0, 
&\qquad 
\tau_{{1}{3}}^{*}(x_{1},H,s) &= 0, \\[4pt]
\tau_{{3}{3}}^{*}(x_{1},0,s) &= 0, 
&\qquad 
\tau_{{3}{3}}^{*}(x_{1},H,s) &= 0, \\[4pt]
D_{{3}}^{*}(x_{1},0,s) &= 0,
&\qquad 
D_{{3}}^{*}(x_{1},H,s) &= 0,
\end{aligned}
\right\}
\qquad 0 \le x_{1} < \infty.
\tag{BC5$^*$}
\label{eq:BC_surfaces_star}
\end{equation}
The general solutions of Eqs.~(\ref{eq:laplace_governing}) can be obtained by employing the Fourier integral transform technique and may be expressed in compact form as
\begin{align}
\mathbf{u}^*(x_1,x_3,s)
=&
\frac{1}{2\pi}\sum_{j=1}^{3}
\int_{-\infty}^{\infty}
\mathbf{N}_{j}(p)\, P_{1j}(p,s)\,
\exp\!\left(|p|\Upsilon_{1j} x_{1}\right)
\exp\!\left(-i p x_{3}\right)\, dp\\
&+
\frac{2}{\pi}\sum_{j=1}^{6}
\int_{0}^{\infty}
\mathbf{M}_{j}(p)\, P_{2j}(p,s)\,
\exp\!\left(p\Upsilon_{2j} x_{3}\right)
\mathbf{F}(p x_{1})\, dp,
\label{eq:phi_general_solution_align}
\end{align}
where
\begin{equation}
\quad
\mathbf{N}_{j}(p) =
\begin{bmatrix}
N_{1j}(p) \\
N_{2j}(p) \\
N_{3j}(p)
\end{bmatrix},
\quad
\mathbf{M}_{j}(p) =
\begin{bmatrix}
M_{1j}(p) \\
M_{2j}(p) \\
M_{3j}(p)
\end{bmatrix},
\quad
\mathbf{F}(p x_{1}) =
\begin{bmatrix}
\sin(p x_{1}) \\
\cos(p x_{1}) \\
\cos(p x_{1})
\end{bmatrix}.
\end{equation}
where $P_{1j}(p,s)$ and $P_{2j}(p,s)$ are unknown spectral functions determined 
from the boundary conditions.
The functions $N_{ij}(p)$ and $M_{ij}(p)$ $(i=1,2,3)$ denote the components of the corresponding eigenvectors, 
while the quantities $\Upsilon_{1j}$ $(j=1,2,3)$ and $\Upsilon_{2j}$ $(j=1,2,\ldots,6)$ represent the characteristic roots obtained from the governing equations. 
The characteristic roots $\Upsilon_{1j}$ and $\Upsilon_{2j}$ together with the associated eigenvectors $N_{ij}$ and $M_{ij}$ are obtained from the homogeneous algebraic systems derived from the transformed governing equations. 
For brevity, the detailed derivation of these algebraic systems is presented in \ref{sec:eigenvectors}.

Substituting the displacement fields 
and electric potential 
\eqref{eq:phi_general_solution_align}
into the constitutive relations, the stress and electric–displacement components of the
piezoelectric strip in Laplace–Fourier transform domain are obtained as
\begin{align}
\boldsymbol{\tau}^*(x_1,x_3,s)
=&
\frac{1}{2\pi}\sum_{j=1}^{3}
\int_{-\infty}^{\infty}
\mathbf{d}_{1j}(p)\, P_{1j}(p,s)\,
\exp\!\left(|p|\Upsilon_{1j} x_{1}\right)
\exp\!\left(-i p x_{3}\right)\, dp\\
&+\frac{2}{\pi}\sum_{j=1}^{6}
\int_{0}^{\infty}
\Big[
\mathbf{d}_{2j}(p)\circ \mathbf{G}(p x_{1})
\Big]\,
P_{2j}(p,s)\,
\exp\!\left(p\Upsilon_{2j} x_{3}\right)\, dp,
\label{eq:Dz}
\end{align}
where
\begin{equation}
\boldsymbol{\tau}^* =
\begin{bmatrix}
\tau^{*}_{11} \\
\tau^{*}_{13} \\
\tau^{*}_{33} \\
D^{*}_{1} \\
D^{*}_{3}
\end{bmatrix},
\end{equation}
and
\begin{equation}
\mathbf{d}_{1j}(p) =
\begin{bmatrix}
p\,d_{11j}(p) \\
|p|\,d_{31j}(p) \\
p\,d_{41j}(p) \\
|p|\,d_{21j}(p) \\
p\,d_{51j}(p)
\end{bmatrix},
\quad
\mathbf{d}_{2j}(p) =
\begin{bmatrix}
p\,d_{12j}(p) \\
p\,d_{32j}(p) \\
p\,d_{42j}(p) \\
p\,d_{22j}(p) \\
p\,d_{52j}(p)
\end{bmatrix},
 \quad
\mathbf{G}(p x_{1}) =
\begin{bmatrix}
\cos(p x_{1}) \\
\sin(p x_{1}) \\
\cos(p x_{1}) \\
\sin(p x_{1}) \\
\cos(p x_{1})
\end{bmatrix}.
\end{equation}
The coefficients $d_{ikj}(p)$ $(i=1,2,\dots,5;\; k=1,2)$ appearing in the stress and electric–displacement components are given in \ref{dikj}. 
To model the crack opening, a displacement discontinuity function $G(x_{3},s)$ is introduced along the crack plane, defined as
\be
\frac{\partial u_{1}^{*}(0,x_{3},s)}{\partial x_{3}} = G(x_{3},s).
\label{G(x3)}
\ee
From the boundary conditions, it follows that the displacement discontinuity vanishes outside the crack region, i.e.,
\be
G(x_{3},s) = 0, \qquad 0 \le x_{3} < a, \;\; b < x_{3} \le H.
\ee
For an embedded crack, the displacement discontinuity must satisfy the single-valued condition
\be
\int_{a}^{b} G(x_{3},s)\,dx_{3} = 0.
\ee
For an edge crack, the corresponding condition reduces to
\be
G(x_{3},s) = 0 \quad \text{at} \quad x_{3} = H.
\ee
From Eq.~(\ref{G(x3)}), substituting the displacement field \eqref{eq:phi_general_solution_align} and evaluating at $x_1=0$, the following relation between the displacement discontinuity and the unknown spectral functions is obtained.
\begin{equation}
\sum_{j=1}^{3}
p\, N_{1j}(p)\, P_{1j}(p,s)
=i \int_{a}^{b} 
G(r,s) e^{i p r}\, dr.
\label{G_transform}
\end{equation}
From the boundary conditions \eqref{eq:BC_tau_xz_star} and \eqref{eq:BC_Dx_star}
\begin{align}
\sum_{j=1}^{3}
\,d_{31j}(p)\,P_{1j}(p,s) &= 0, \\
\sum_{j=1}^{3}
\,d_{21j}(p)\,P_{1j}(p,s) &= 0.
\label{system P1j}
\end{align}
By solving Eqs.~\eqref{G_transform} and \eqref{system P1j}, the unknown spectral functions $P_{1j}(p,s)$ are obtained as follows:
\begin{equation}
P_{1j}(p,s)
=
\frac{i\,\lambda_j(p)}{p}
\int_{a}^{b}
G(r,s)\,e^{ipr}\,dr, 
\qquad j=1,2,3.
\label{P1_solution}
\end{equation}
The auxiliary constants $\lambda_j(p)$ are defined as
\begin{align}
\lambda_1(p) &= \frac{d_{312}(p)d_{213}(p)-d_{313}(p)d_{212}(p)}{\Delta(p)},\\[4pt]
\lambda_2(p) &= \frac{d_{211}(p)d_{313}(p)-d_{311}(p)d_{213}(p)}{\Delta(p)},\\[4pt]
\lambda_3(p) &= \frac{d_{311}(p)d_{212}(p)-d_{211}(p)d_{312}(p)}{\Delta(p)},
\end{align}
where the determinant $\Delta(p)$ is given by
\begin{align}
\Delta(p) &= N_{11}(p)\big(d_{312}(p)d_{213}(p)-d_{313}(p)d_{212}(p)\big)+ N_{12}(p)\big(d_{211}(p)d_{313}(p)-d_{311}(p)d_{213}(p)\big)\nonumber\\
&\quad + N_{13}(p)\big(d_{311}(p)d_{212}(p)-d_{211}(p)d_{312}(p)\big).
\end{align} 
With the help of the boundary conditions \eqref{eq:BC_surfaces_star} and Eq.~\eqref{P1_solution}, 
the unknown spectral functions $P_{2j}(p,s)$ can be obtained as
\begin{equation}
P_{2j}(p,s)
=
\frac{i}{2\pi p}
\int_{a}^{b}
\psi_j(p,r)\,G(r,s)\,dr, 
\qquad j=1,2,\ldots,6,
\label{P2_solution}
\end{equation}
where $\psi_j(p,r)$ are defined in \ref{matrix theta}.

Now, by employing the boundary condition \eqref{eq:BC_tau_xx_star}, 
the singular integral equation governing the unknown function $G(x_3,s)$ is obtained:
\begin{equation}
\frac{\eta_0}{\pi}\int_a^b
\frac{G(r,s)}{r-x_3}\,dr
+
\int_a^b
H_1(x_3,r,s)\,G(r,s)\,dr
=
-\frac{\tau_{0}}{s} - \tau_{0}^{T*}(x_{3},s),
\qquad a<x_3<b ,
\label{eq:SIE_final}
\end{equation}
The regular kernel $H_1(x_3,r,s)$ is given by
\begin{equation}
H_1(x_3,r,s)
=
\frac{i}{\pi^2}
\int_{0}^{\infty}
\Lambda_1(s,x_3,r,p)\,dp,
\label{eq:H1_def}
\end{equation}
with
\begin{equation}
\Lambda_1(s,x_3,r,p)
=
\sum_{j=1}^{6}
d_{12j}(p)\,e^{p\Upsilon_{2j} x_3}\,\psi_j(p,r),
\label{eq:Lambda1_def}
\end{equation}
and
\begin{equation}
\eta_0 = -\frac{|p|}{p}\sum_{j=1}^{3} d_{11j}\,\lambda_j(p).
\end{equation}
To transform the finite interval $[a,b]$ into the standard interval $[-1,1]$, define linear mapping
\begin{equation}
r=\frac{b-a}{2}\,\varsigma+\frac{b+a}{2},
\qquad
x_3=\frac{b-a}{2}\,\varsigma_0+\frac{b+a}{2},
\qquad \varsigma,\varsigma_0 \in [-1,1].
\end{equation}
Accordingly, the kernel transforms to
\begin{equation}
\widetilde H_1(\varsigma,\varsigma_0,s)
=
\frac{i}{\pi^2}
\int_{0}^{\infty}
\Lambda_1
\!\left(
s,
\frac{b-a}{2}\varsigma_0+\frac{b+a}{2},
\frac{b-a}{2}\varsigma+\frac{b+a}{2},
p
\right)\,dp .
\end{equation}
Under this transformation, Eq.~\eqref{eq:SIE_final} reduces to the standard Cauchy-type singular integral equation
\begin{equation}
\frac{\eta_0}{\pi}
\int_{-1}^{1}
\frac{\widetilde G(\varsigma,s)}{\varsigma-\varsigma_0}\,d\varsigma+\frac{b-a}{2}\int_{-1}^{1}
\widetilde H_1(\varsigma,\varsigma_0,s)\,
\widetilde G(\varsigma,s)\,d\varsigma=-\frac{\tau_{0}}{s}-\widetilde{\tau}_0^{T}(\varsigma_0,s),
\quad -1<\varsigma_0<1.
\label{eq:standard_final}
\end{equation}
Here, the transformed quantities are defined as
\begin{equation}
\widetilde G(\varsigma,s)
=
G\!\left(\frac{b-a}{2}\varsigma+\frac{b+a}{2},\,s\right),
\qquad
\widetilde{\tau}_0^{T}(\varsigma_0,s)
=
\tau_0^{T*}\!\left(\frac{b-a}{2}\varsigma_0+\frac{b+a}{2},\,s\right).
\end{equation}
To obtain the numerical solution of the singular integral equation~\eqref{eq:standard_final}, 
the Lobatto-Chebyshev collocation technique is employed.
Since the solution exhibits square-root singular behaviour at the crack tips, 
the unknown function is expressed in the weighted form
\begin{equation}
\Psi(\varsigma,s)
=
\sqrt{1-\varsigma^{2}}\,\widetilde G(\varsigma,s),
\qquad |\varsigma|\le1 .
\label{eq:weighted_unknown}
\end{equation}
The Lobatto-Chebyshev quadrature nodes are chosen as
\begin{equation}
\varsigma_k=\cos\!\left(
\frac{(k-1)\pi}{N-1}
\right),
\qquad k=1,2,\ldots,N ,
\end{equation}
while the collocation points are
\begin{equation}
\varsigma_j^{(0)}
=
\cos\!\left(
\frac{(2j-1)\pi}{2(N-1)}
\right),
\qquad j=1,2,\ldots,N-1 .
\end{equation}
The associated Lobatto weights are given by
\[
w_k=\frac{\pi}{2(N-1)}, \quad k=1,N,\qquad
w_k=\frac{\pi}{N-1}, \quad k=2,3,\ldots,N-1.
\]
Applying the Lobatto-Chebyshev method \cite{yang2022dynamic} to Eq.~\eqref{eq:standard_final}, 
the following system of linear algebraic equations is obtained:
\begin{equation}
\sum_{k=1}^{N}
w_k\,\Psi(\varsigma_k,s)
\left[
\frac{\eta_0}{\pi}
\frac{1}{\varsigma_k-\varsigma_j^{(0)}}
+
\frac{b-a}{2}\,
\widetilde H_1(\varsigma_k,\varsigma_j^{(0)},s)
\right]
=
-\frac{\tau_0}{s}
-
\widetilde{\tau}_0^{T}(\varsigma_j^{(0)},s),
\quad j=1,2,\ldots,N-1 .
\label{eq:algebraic_system}
\end{equation}
In addition, the subsidiary condition is written as
\begin{equation}
\sum_{k=1}^{N}
w_k\,\Psi(\varsigma_k,s)=0 .
\label{eq:subs_discrete}
\end{equation}
Therefore, the stress intensity factors $K_{Ia}^{\ast}(s)$ and $K_{Ib}^{\ast}(s)$ in the Laplace domain are given by \cite{wang2002cracked}
\begin{equation}
K^{\ast}_{Ia}(s)
=
\lim_{x_{3} \to a^{-}}
\left\{ 
\sqrt{2\pi(a - x_{3})}\;
\tau^{\ast}_{11}(0,x_{3},s)
\right\}
=
\eta_0\, \sqrt{\pi c}\; \Psi(-1,s),
\label{eq:KIa}
\end{equation}
\begin{equation}
K^{\ast}_{Ib}(s)
=
\lim_{x_{3} \to b^{+}}
\left\{ 
\sqrt{2\pi(x_{3} - b)}\;
\tau^{\ast}_{11}(0,x_{3},s)
\right\}
=
-\eta_0\, \sqrt{\pi c}\; \Psi(1,s),
\label{eq:KIb}
\end{equation}
These expressions represent the fundamental fracture parameters of the problem and constitute the primary quantities of interest in the present study. In the subsequent sections, these results will be evaluated numerically and used to analyze the effects of thermal shock, fractional parameters, and pre-existing stresses on crack behavior through graphical illustrations.

\section{Numerical Results and Discussion}
\label{Numerical Results and Discussion}
\subsection{Numerical Inversion of Laplace Transform}
\label{Numerical Inversion of Laplace Transform}
To obtain the temperature field and the dynamic thermal stress intensity factors in the time domain, it is necessary to perform the inverse Laplace transform of the corresponding quantities derived in the Laplace domain. However, due to the complexity of the governing equations, an explicit analytical inversion is not feasible. Therefore, a numerical inversion technique is employed.
In the present study, the numerical inversion scheme proposed by Stehfest~\cite{stehfest1970algorithm} is adopted to evaluate the time-dependent response. Let $\Xi^{\ast}(1,s)$ represent a function in the Laplace domain. Its inverse Laplace transform $\Xi(1,t)$ can be approximated as
\begin{equation}
\Xi(1,t) = \frac{\ln 2}{t} \sum_{n=1}^{2L} V_n , \Xi^{\ast}\left(1, \frac{n \ln 2}{t}\right),
\end{equation}
where $L$ is a positive integer controlling the accuracy of the inversion, and $V_n$ are the Stehfest weighting coefficients defined by
\begin{equation}
V_n = (-1)^{n+L} \sum_{m=\left\lfloor \frac{n+1}{2} \right\rfloor}^{\min(n,L)}
\frac{m^L (2m)!}{(L-m)!  m!  (m-1)!  (n-m)! (2m-n)!}.
\end{equation}
Here, $\lfloor \cdot \rfloor$ denotes the floor function, i.e., the greatest integer less than or equal to the argument. The parameter $L$ is typically chosen as a small even integer (e.g., $L = 6, 8,$ or $10$) to ensure numerical stability and accuracy.
The Stehfest method is advantageous due to its simplicity, as it involves only a single parameter, unlike other inversion techniques that require multiple tuning parameters. Using this approach, the temperature field and the corresponding stress intensity factors are efficiently evaluated in the time domain from their Laplace-domain representations.

\subsection{Material Properties and Numerical Parameters}
\label{Material Properties and Numerical Parameters}
Numerical simulations are performed using MATLAB to evaluate the temperature distribution, thermoelastic stress fields, and thermal stress intensity factors at the crack tips located at $x_3=a$ and $x_3=b$. The analysis examines the effects of fractional order, thermal relaxation time, Fourier number, initial stresses, and crack length on the thermoelastic response of the medium.
The material properties of PZT-4 employed in this study are summarized in Table~\ref{tab:material}. These parameters are adopted to generate the numerical results and ensure consistency with established piezoelectric material data.
\begin{table}[htbp]
\centering
\caption{Material constants for PZT-4 \cite{zhou2018moving}}
\label{tab:material}
\renewcommand{\arraystretch}{1.5}
\begin{tabular}{llll}
\hline
\textbf{Parameter} & \textbf{Value} & \textbf{Parameter} & \textbf{Value} \\
\hline

$\mu_{11}$ (N/m$^2$) & $139 \times 10^{9}$ 
& $\kappa_{11}$ (N/(K\,m$^2$)) & $0.973 \times 10^{6}$ \\

$\mu_{13}$ (N/m$^2$) & $74.3 \times 10^{9}$ 
& $\kappa_{33}$ (N/(K\,m$^2$)) & $0.791 \times 10^{6}$ \\

$\mu_{33}$ (N/m$^2$) & $113 \times 10^{9}$ 
& $\varepsilon_{11}$ (C/(V\,m)) & $60 \times 10^{-10}$ \\

$\mu_{44}$ (N/m$^2$) & $25.6 \times 10^{9}$ 
& $\varepsilon_{33}$ (C/(V\,m)) & $54.7 \times 10^{-10}$ \\

$e_{31}$ (C/m$^2$) & $-6.98$ 
& $p_{z}$ (C/(K\,m$^2$)) & $-48.86 \times 10^{-6}$ \\

$e_{33}$ (C/m$^2$) & $13.84$ 
& $\rho$ (kg/m$^3$) & $7.5 \times 10^{3}$ \\

$e_{15}$ (C/m$^2$) & $-13.44$ 
& $k_{1}$ (W/(K\,m)) & $0.52$ \\
$e_{32}$ (C/m$^2$) & $-6.15$ 
& $k_{3}$ (W/(K\,m)) & $0.12$ \\
\hline
\end{tabular}
\end{table}

\subsection{Temperature and Electromechanical Distribution in the Uncracked Strip}
\label{Temperature and Electromechanical Distribution in the Uncracked Strip}
\begin{figure}[htbp]
\centering
\begin{subfigure}[b]{0.47\textwidth}
\centering
\includegraphics[width=\textwidth]{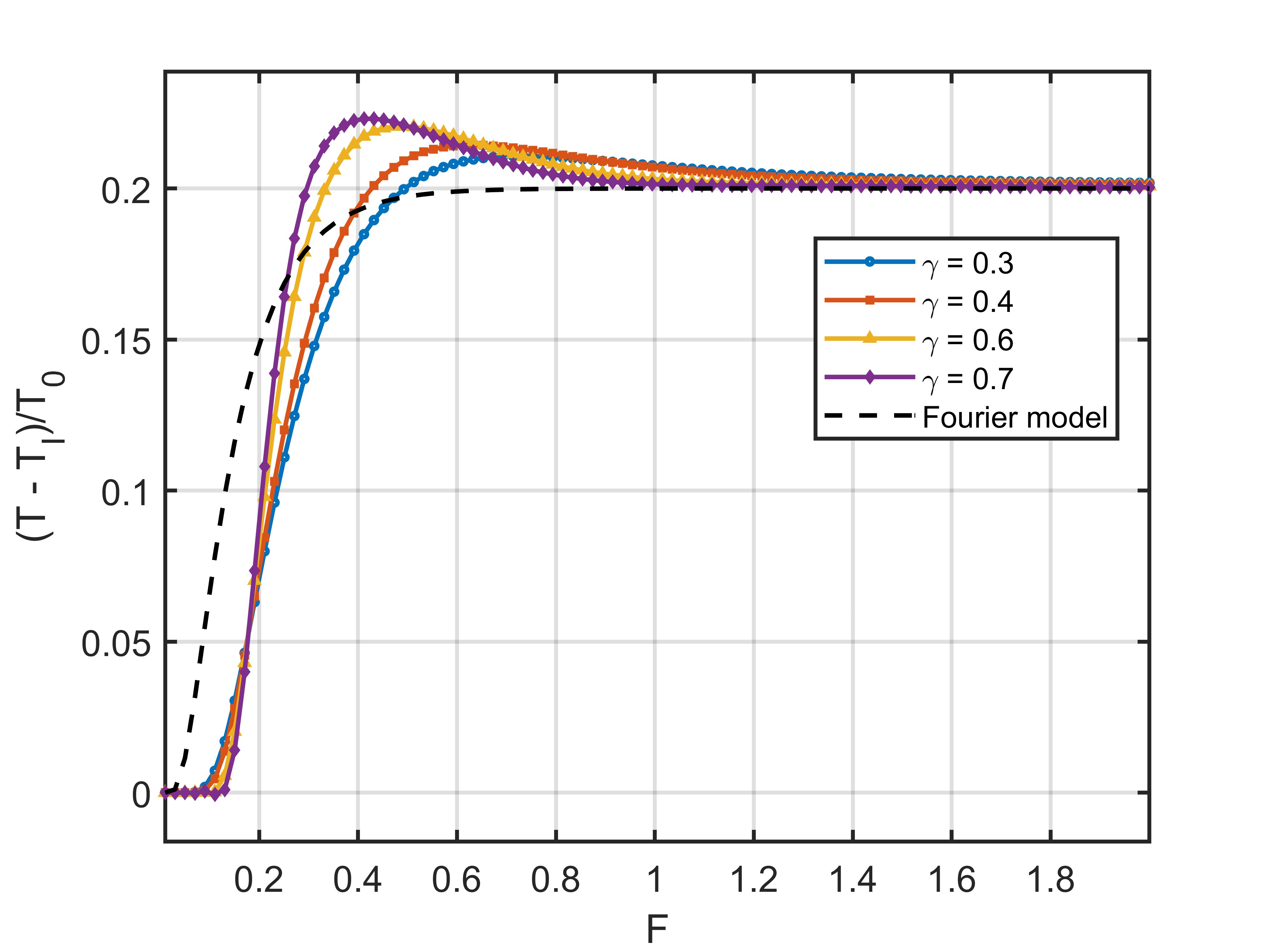}
\caption{Effect of fractional order $\gamma$}
\label{fig:2a}
\end{subfigure}
\hfill
\begin{subfigure}[b]{0.47\textwidth}
\centering
\includegraphics[width=\textwidth]{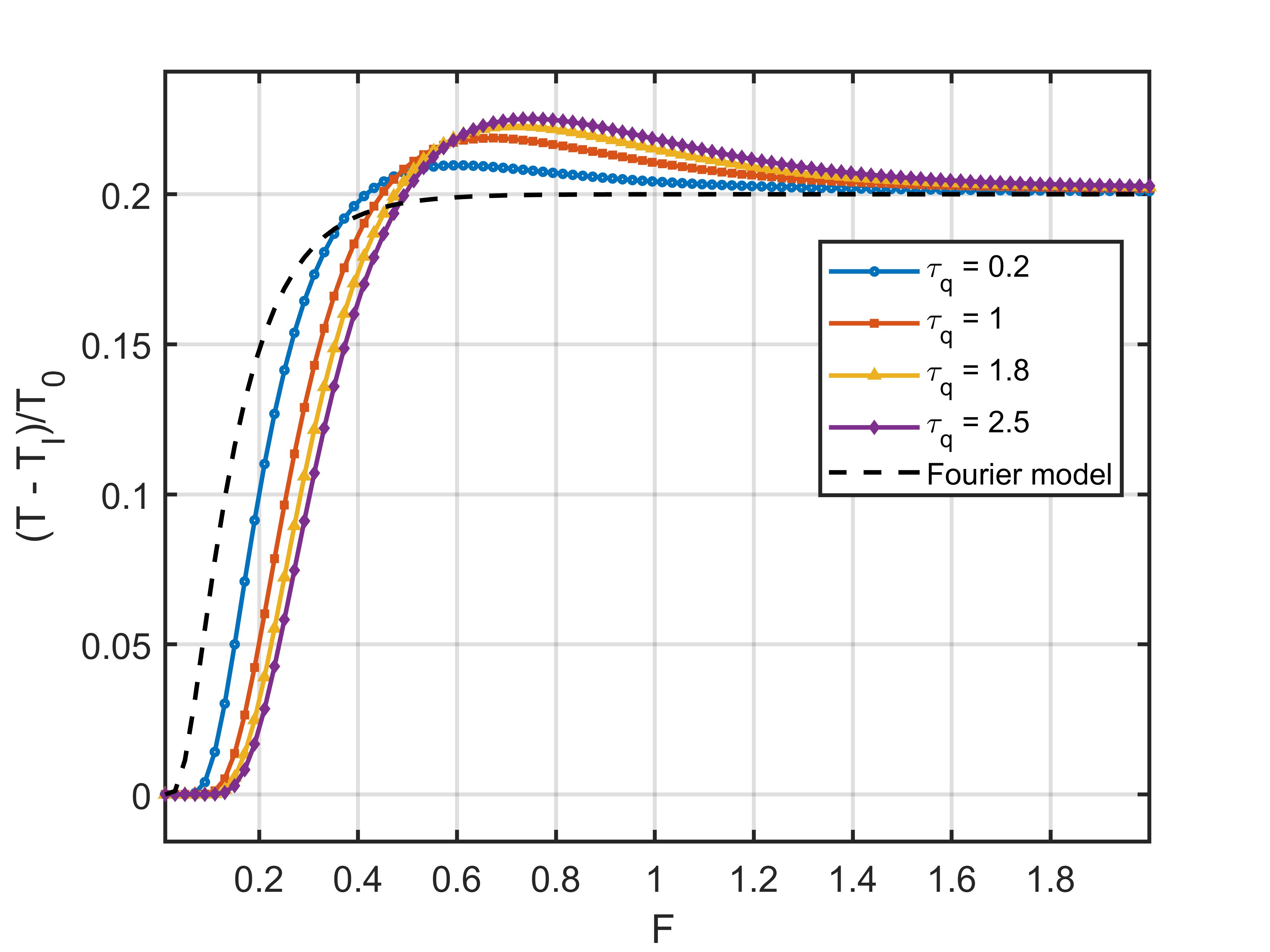}
\caption{Effect of relaxation time $\tau_q$}
\label{fig:2b}
\end{subfigure}
\vspace{0.4cm}
\begin{subfigure}[b]{0.47\textwidth}
\centering
\includegraphics[width=\textwidth]{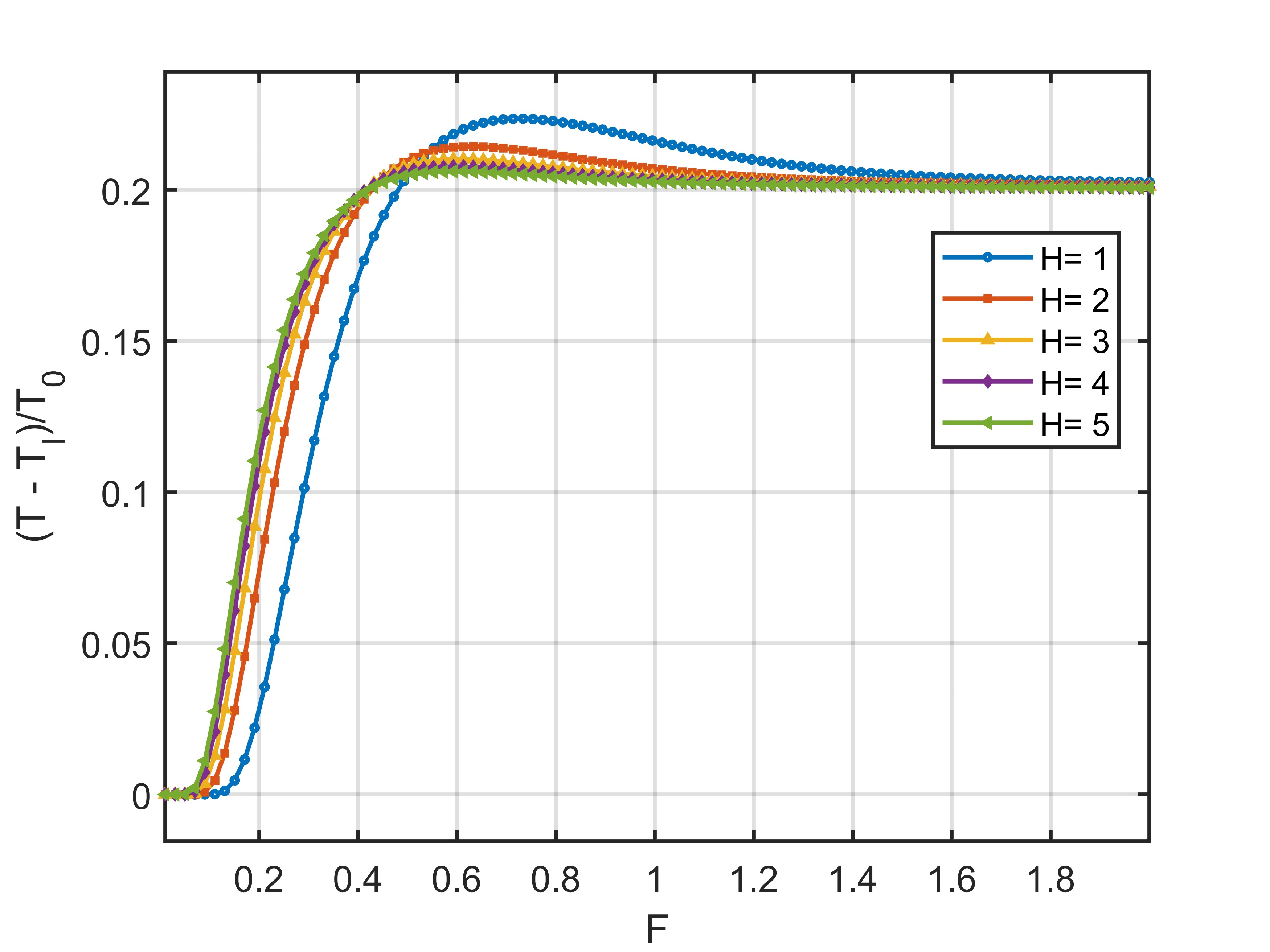}
\caption{Effect of strip thickness $H$}
\label{fig:2c}
\end{subfigure}
\hfill
\begin{subfigure}[b]{0.47\textwidth}
\centering
\includegraphics[width=\textwidth]{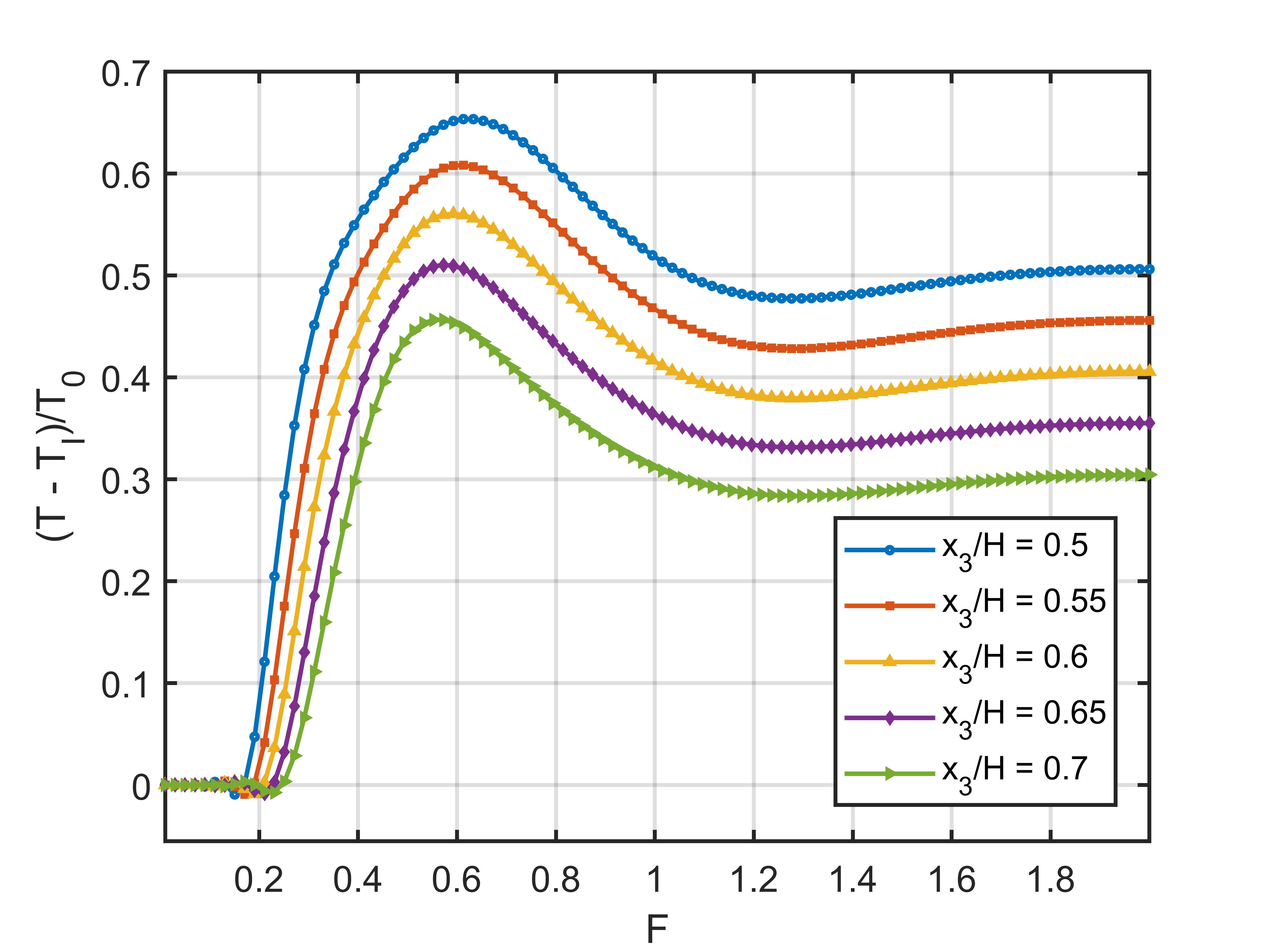}
\caption{Effect of normalized depth $x_3/H$}
\label{fig:2d}
\end{subfigure}
\caption{Variation of nondimensional temperature with Fourier number for different values of fractional order $\gamma$, thermal relaxation time $\tau_q$, strip thickness $H$, and normalized depth $x_3/H$.}
\label{fig:2}
\end{figure}

Figure~\ref{fig:2} illustrates the variation of nondimensional temperature with respect to Fourier number $F = \frac{\lambda_0 t}{H^2}$ for different physical parameters, namely the fractional order $\gamma$, thermal relaxation time $\tau_q$, strip thickness $H$, and normalized depth $x_3/H$.
Figure~\ref{fig:2a} shows that the temperature increases rapidly with the Fourier number and asymptotically approaches a steady-state. As the fractional order $\gamma$ increases, the temperature response becomes faster and reaches the steady state more quickly. This indicates that higher fractional order reduces the memory effect and enhances heat propagation within the medium.
Figure~\ref{fig:2b} depicts the influence of thermal relaxation time $\tau_q$. It is noted that increasing $\tau_q$ delays the thermal response, resulting in a slower rise in temperature. This behavior reflects the non-Fourier heat conduction effect, where thermal disturbances propagate with finite speed.
The effect of strip thickness $H$ is presented in Fig.~\ref{fig:2c}. It is observed that an increase in thickness slightly reduces the peak temperature and smoothens the temperature distribution. This is due to the increased domain over which heat diffusion occurs.
Figure~\ref{fig:2d} shows the variation of temperature with the normalized depth $x_3/H$. It is observed that the temperature is maximum near the heated boundary and decreases progressively with increasing depth, indicating attenuation of thermal energy as it propagates into the strip.
The results, when compared with the classical Fourier model, show that the fractional formulation yields a delayed and smoother thermal response due to the presence of memory effects in heat conduction.

\begin{figure}[htbp]
\centering

\begin{subfigure}[b]{0.47\textwidth}
\centering
\includegraphics[width=\textwidth]{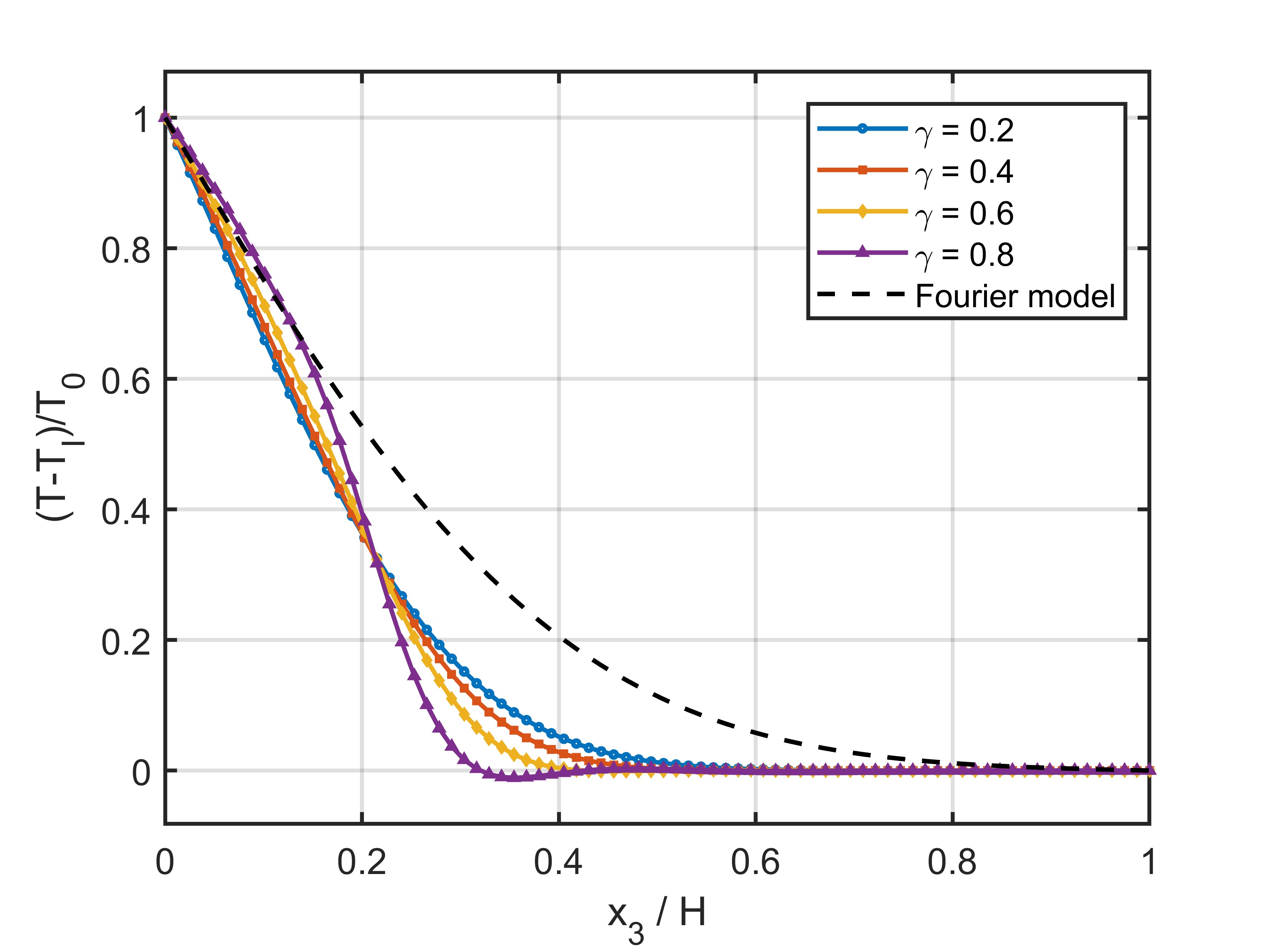}
\caption{Effect of fractional order $\gamma$}
\label{fig:3a}
\end{subfigure}
\hfill
\begin{subfigure}[b]{0.47\textwidth}
\centering
\includegraphics[width=\textwidth]{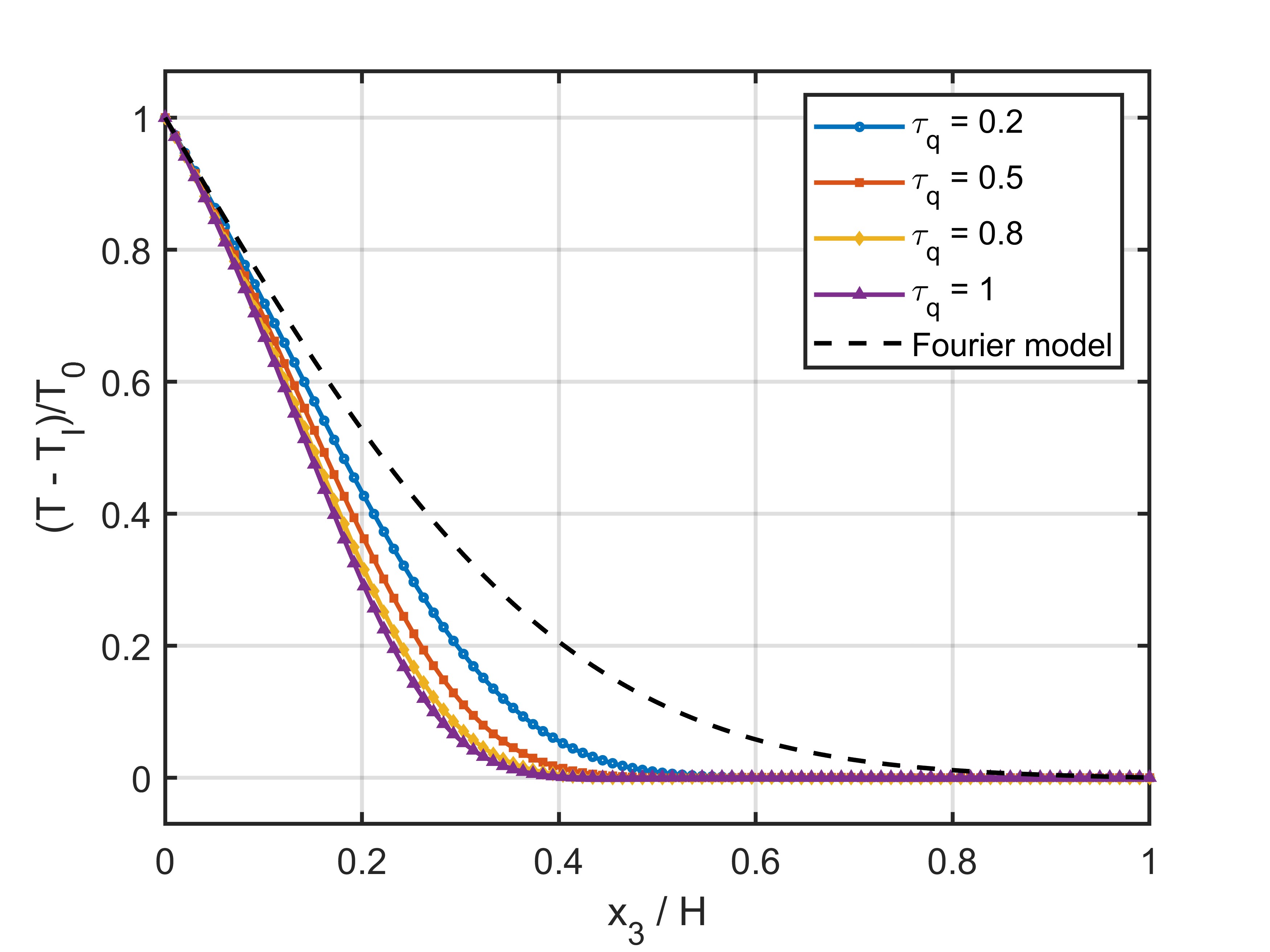}
\caption{Effect of relaxation time $\tau_q$}
\label{fig:3b}
\end{subfigure}

\vspace{0.4cm}

\begin{subfigure}[b]{0.47\textwidth}
\centering
\includegraphics[width=\textwidth]{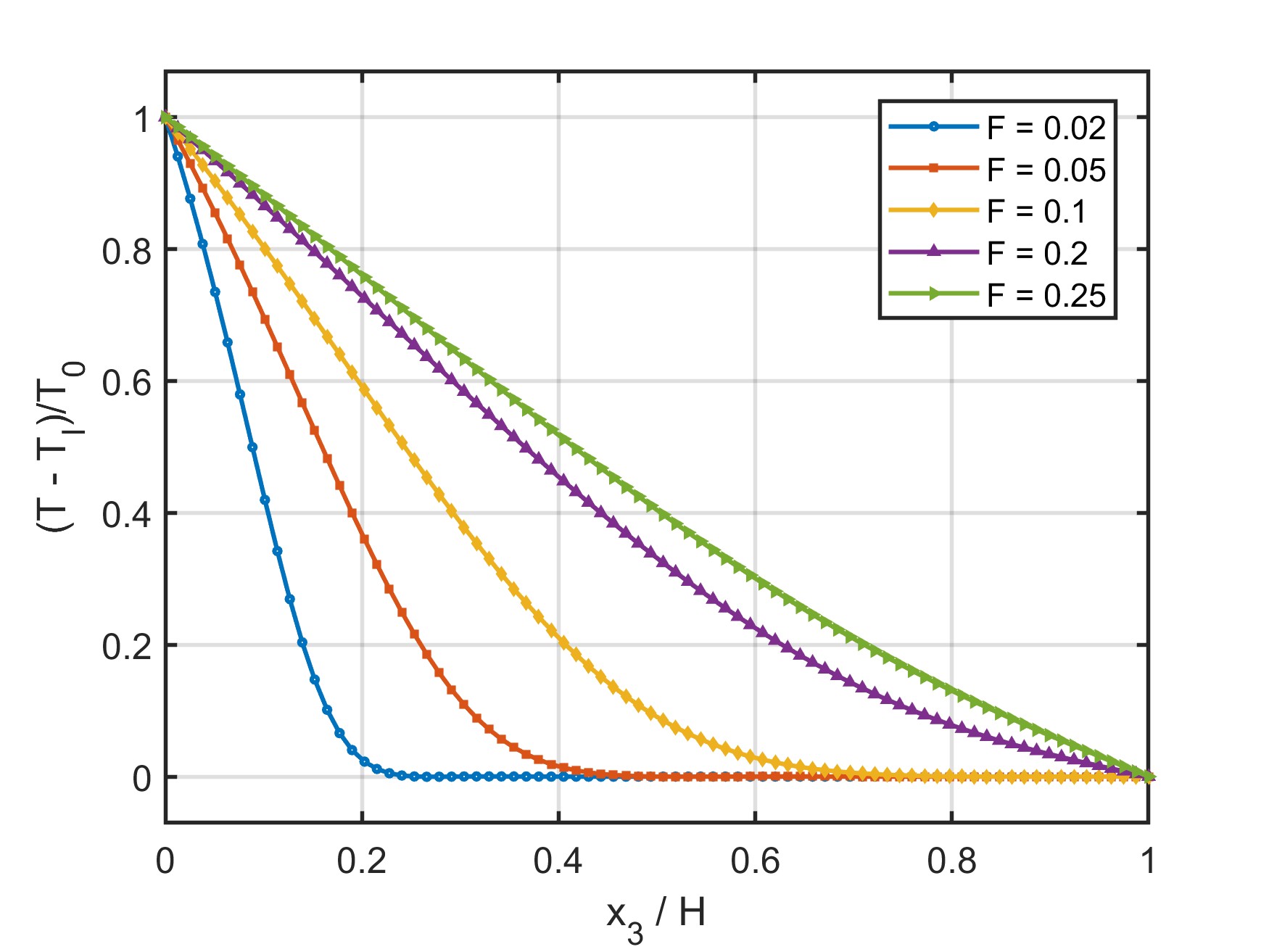}
\caption{Effect of Fourier number $F$}
\label{fig:3c}
\end{subfigure}
\hfill
\begin{subfigure}[b]{0.47\textwidth}
\centering
\includegraphics[width=\textwidth]{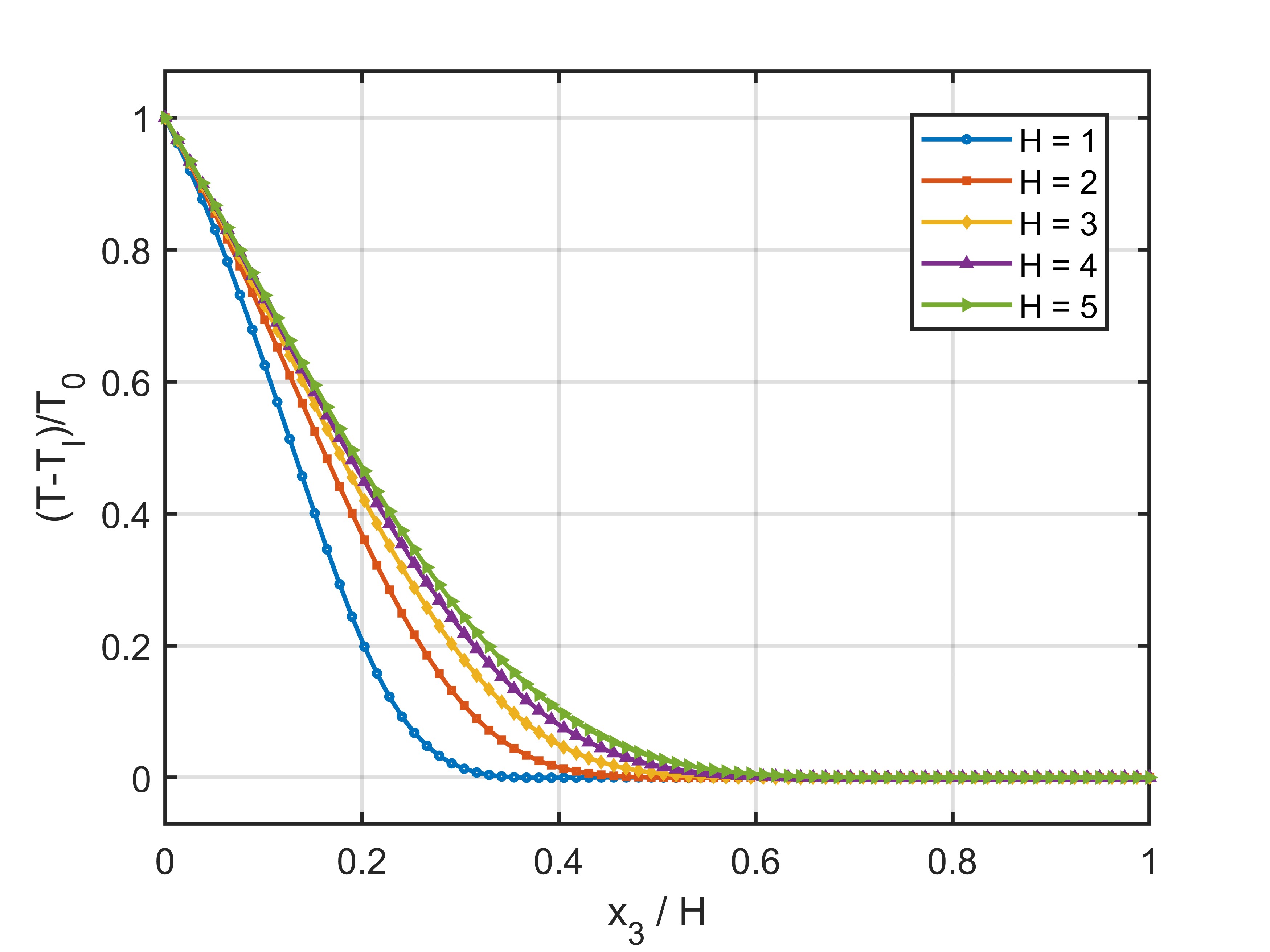}
\caption{Effect of strip thickness $H$}
\label{fig:3d}
\end{subfigure}

\caption{Variation of nondimensional temperature with normalized depth $x_3/H$ for different values of fractional order $\gamma$, thermal relaxation time $\tau_q$, Fourier number $F$, and strip thickness $H$.}
\label{fig:3}

\end{figure}

Figure~\ref{fig:3} presents the variation of nondimensional temperature with normalized depth $x_3/H$ for different values of fractional order $\gamma$, thermal relaxation time $\tau_q$, Fourier number $F$, and strip thickness $H$.
Figure~\ref{fig:3a} shows that the temperature decreases rapidly with increasing 
$x_3/H$, indicating attenuation of thermal energy away from the heated boundary. As the fractional order $\gamma$ increases, the temperature decays more rapidly, suggesting that higher fractional order enhances the rate of heat dissipation within the medium.
Figure~\ref{fig:3b} illustrates the effect of thermal relaxation time $\tau_q$. It is observed that increasing $\tau_q$ results in a faster decay of temperature, indicating enhanced heat dissipation and a stronger deviation from classical Fourier behavior.
The influence of the Fourier number $F$ is depicted in Fig.~\ref{fig:3c}. It is observed that higher values of $F$ result in a more gradual temperature decay, implying that the system approaches thermal equilibrium over a longer spatial extent.
Figure~\ref{fig:3d} shows the effect of strip thickness $H$. It is evident that increasing the thickness results in a smoother temperature distribution, with a slower reduction in temperature along the depth, due to the increased domain available for heat conduction.

\begin{figure}[htbp]
\centering
\begin{subfigure}[b]{0.47\textwidth}
\centering
\includegraphics[width=\textwidth]{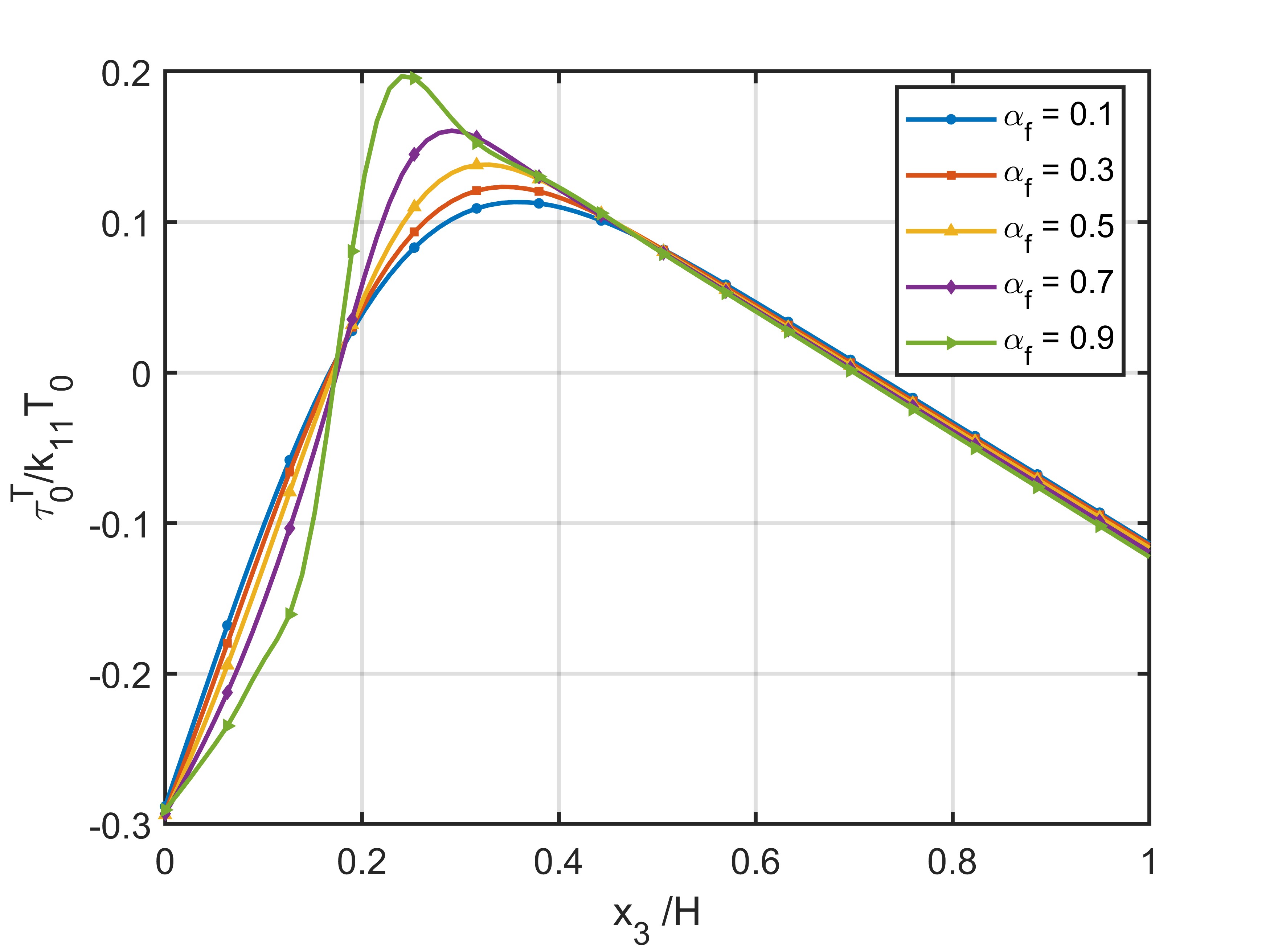}
\caption{Effect of fractional order $\gamma$}
\label{fig:4a}
\end{subfigure}
\hfill
\begin{subfigure}[b]{0.47\textwidth}
\centering
\includegraphics[width=\textwidth]{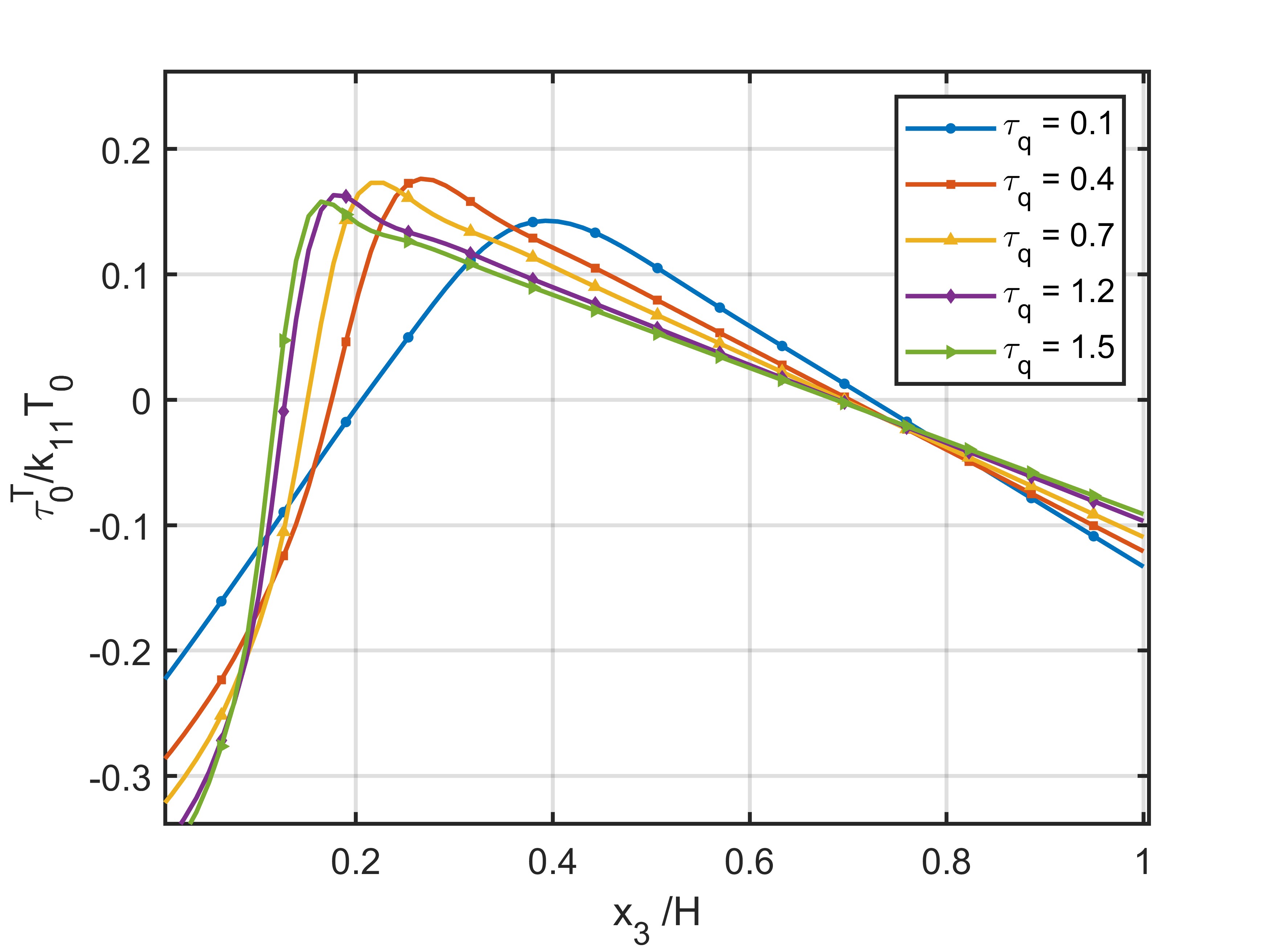}
\caption{Effect of relaxation time $\tau_q$}
\label{fig:4b}
\end{subfigure}

\vspace{0.4cm}

\begin{subfigure}[b]{0.47\textwidth}
\centering
\includegraphics[width=\textwidth]{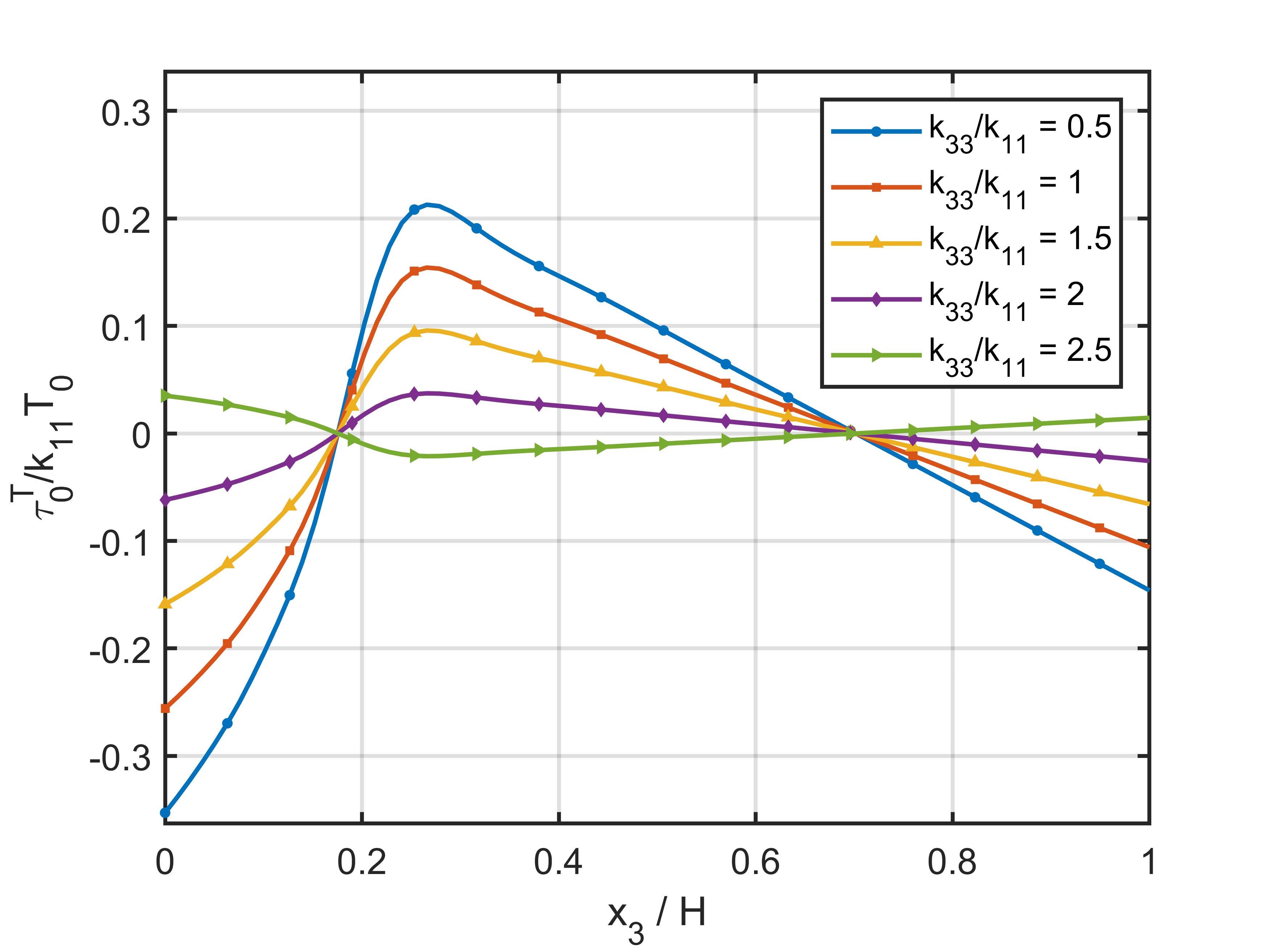}
\caption{Effect of thermal moduli ratio $k_{33}/k_{11}$}
\label{fig:4c}
\end{subfigure}
\hfill
\begin{subfigure}[b]{0.47\textwidth}
\centering
\includegraphics[width=\textwidth]{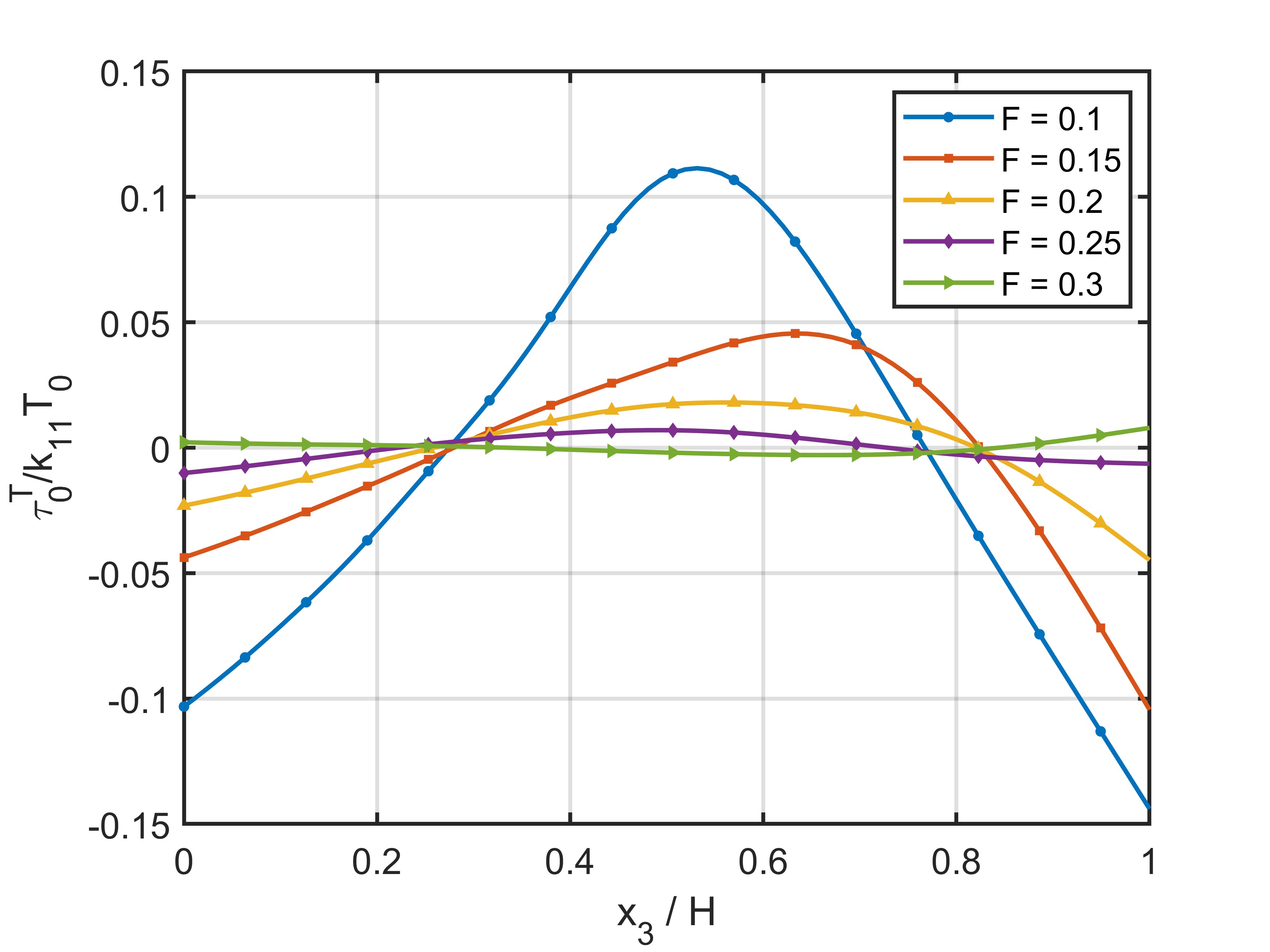}
\caption{Effect of Fourier number $F$}
\label{fig:4d}
\end{subfigure}

\caption{Variation of nondimensional thermal stress $\tau_0^T/(k_{11}T_0)$ with normalized depth $x_3/H$ for different values of fractional order $\gamma$, thermal relaxation time $\tau_q$, thermal moduli ratio $k_{33}/k_{11}$, and Fourier number $F$.}
\label{fig:4}
\end{figure}

Figure~\ref{fig:4} illustrates the variation of the nondimensional thermal stress $\tau_0^T/(k_{11}T_0)$ with the normalized depth $x_3/H$ for different values of fractional order $\gamma$, thermal relaxation time $\tau_q$, thermal modulus ratio $k_{33}/k_{11}$, and Fourier number $F$.
From Fig.~\ref{fig:4a} it is observed that the thermal stress initially increases from a compressive state near the boundary, attains a maximum tensile value, and then gradually decreases along the depth. As the fractional order $\gamma$ increases, the peak stress becomes higher and shifts slightly towards the boundary, indicating a stronger thermoelastic response due to reduced memory effects.
Figure~\ref{fig:4b} shows the influence of thermal relaxation time $\tau_q$. It is observed that increasing $\tau_q$ shifts the peak thermal stress toward the boundary (smaller values of $x_3/H$), indicating reduced penetration of thermal disturbances due to thermal relaxation effects. The peak magnitude exhibits slight variation without a consistent trend.
The effect of the thermal modulus ratio $k_{33}/k_{11}$ is presented in Fig.~\ref{fig:4c}.It is observed that increasing this ratio reduces the peak thermal stress and modifies the stress distribution along the depth, indicating a strong influence of anisotropic thermal properties on the stress field.
Figure~\ref{fig:4d} depicts the variation with the Fourier number $F$. It is evident that for smaller values of $F$, the stress variation is more pronounced, while increasing $F$ results in a smoother distribution and reduced stress gradients, indicating progression towards thermal equilibrium.

\begin{figure}
    \centering
    \includegraphics[width=0.8\linewidth]{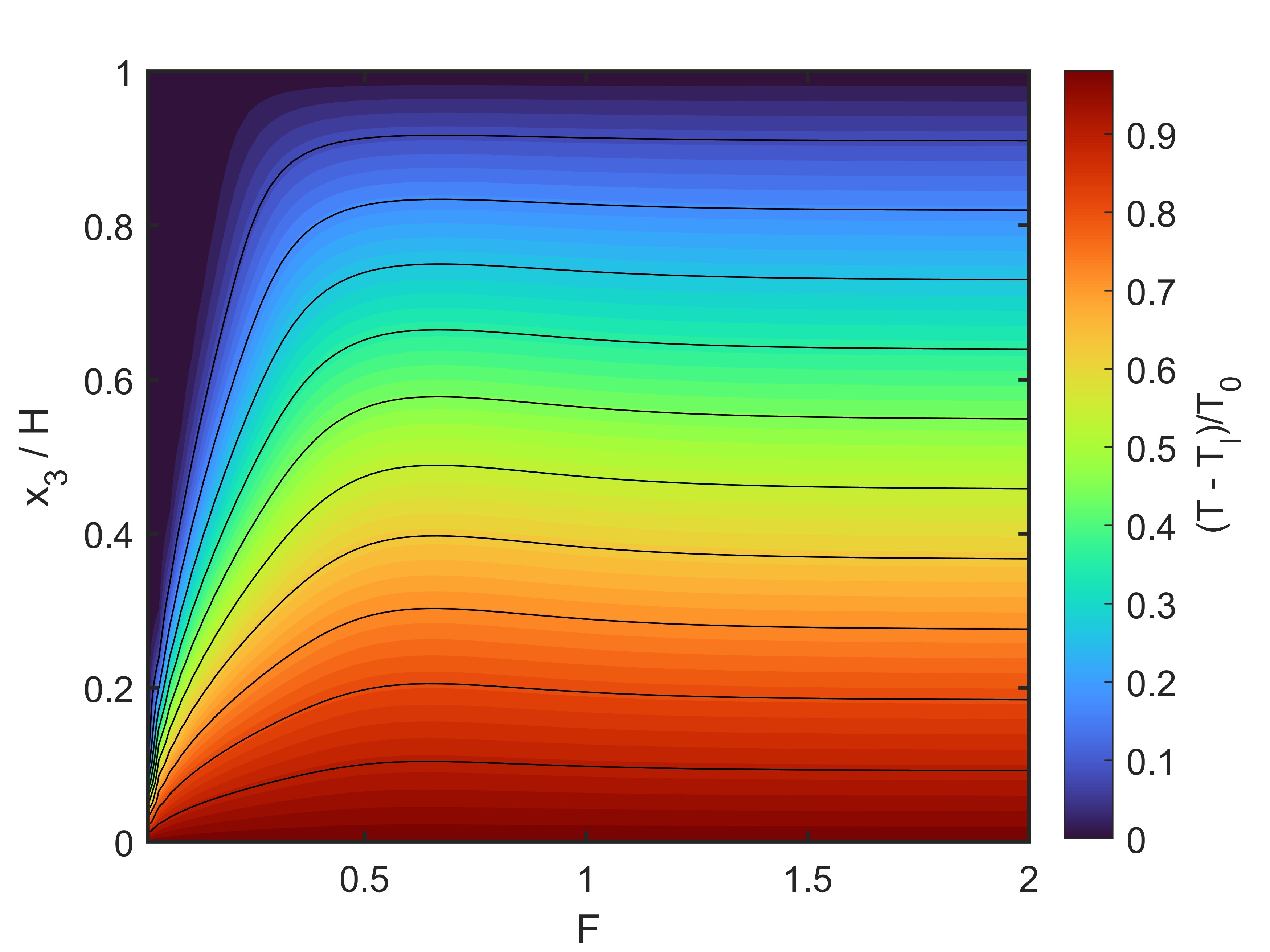}
    \caption{Contour plot of the normalized temperature distribution $(T - T_I)/T_0$ as a function of the Fourier number $F$ and normalized thickness $x_3/H$, showing transient heat propagation and gradual transition to steady state.}
    \label{surf_temp}
\end{figure}

The contour plot in Fig.~\ref{surf_temp} presents the spatial and temporal evolution of the normalized temperature distribution $(T - T_I)/T_0$ as a function of the Fourier number $F$ and the normalized thickness coordinate $x_3/H$. It is observed that at early times (small values of $F$), the temperature rise is highly concentrated near the heated boundary ($x_3/H \approx 0$), resulting in steep thermal gradients and limited heat penetration into the medium. As time progresses, the thermal field propagates deeper into the material, and the temperature distribution gradually becomes smoother, indicating the diffusion of heat across the thickness.
The contour lines reveal a rapid initial thermal response followed by a slower evolution, which is characteristic of transient heat conduction. For larger values of $F$, the temperature field approaches a quasi-steady state, where the variation across the thickness becomes minimal and the contours tend to flatten.
This behavior highlights the influence of fractional heat conduction, where memory effects introduce a delay in thermal propagation compared to classical Fourier models. As a result, the temperature evolution exhibits a non-local and history-dependent behavior, leading to a gradual transition from the transient regime to thermal equilibrium.

\begin{figure}
    \centering
    \includegraphics[width=0.8\linewidth]{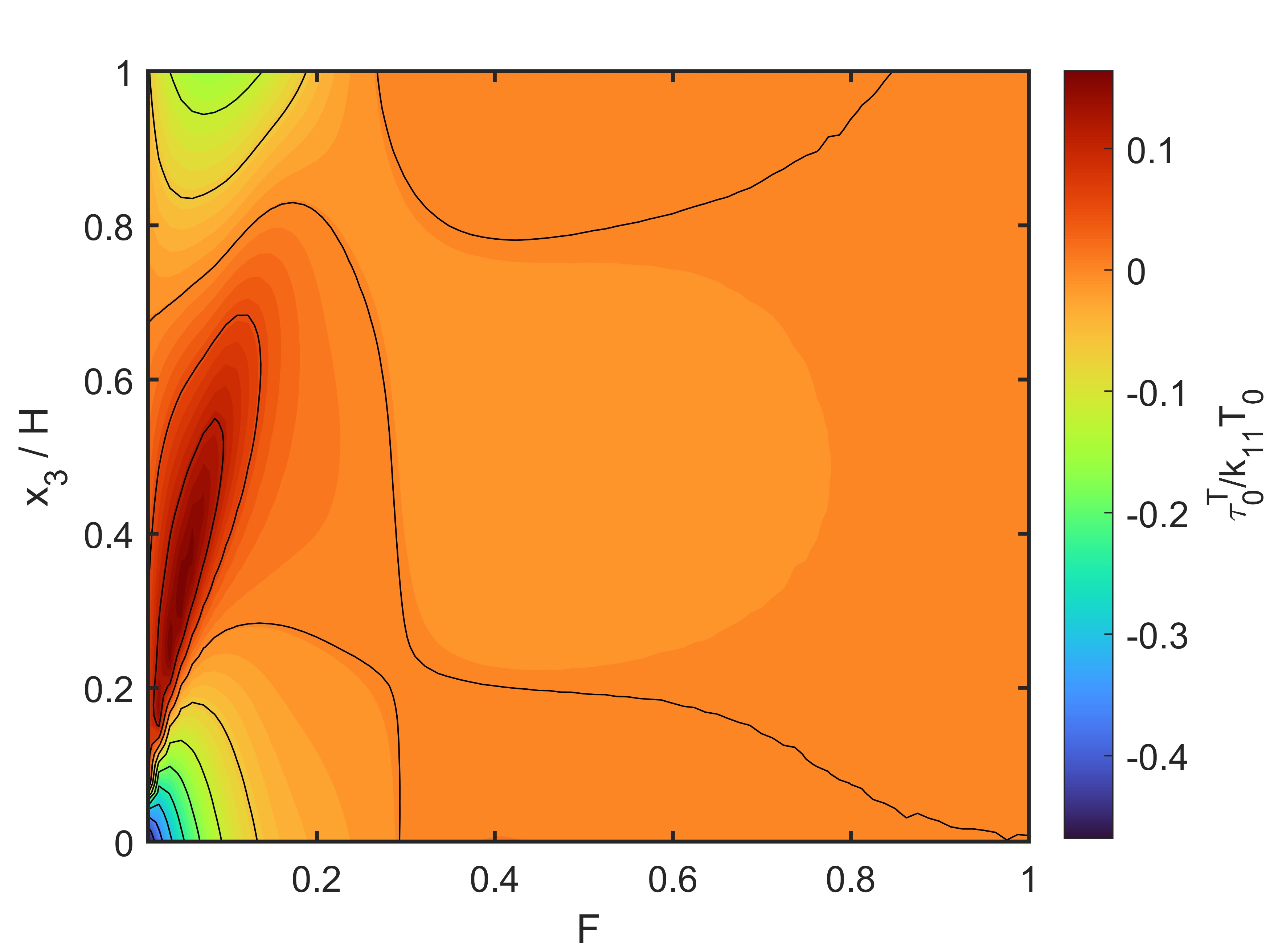}
   \caption{Contour plot of the normalized thermal stress $\tau_0^T/(k_{11}T_0)$ as a function of the Fourier number $F$ and normalized thickness $x_3/H$, showing stress concentration near the heated boundary, stress reversal, and gradual relaxation toward steady state.}
    \label{surf_sigma}
\end{figure}

The contour plot shown in Fig.~\ref{surf_sigma} illustrates the spatial and temporal variation of the normalized thermal stress $\tau/(k_{11}T_0)$ as a function of the Fourier number $F$ and the normalized thickness coordinate $x_3/H$. At early times (small values of $F$), significant stress concentrations are observed near the heated boundary ($x_3/H \approx 0$), where sharp thermal gradients induce strong compressive stresses. Slightly away from the boundary, a transition to tensile stress is evident, indicating the presence of stress reversal due to constrained thermal expansion.
As time increases, the thermal disturbance propagates into the medium, and the stress field gradually redistributes across the thickness. The contour lines reveal that the magnitude of stress decreases and becomes more uniform with increasing $F$, reflecting the relaxation of thermal gradients. At larger values of $F$, the stress distribution approaches a quasi-steady state, where variations are minimal and the material tends toward equilibrium.
The observed behavior highlights the coupled thermoelastic response of the material under transient heating conditions. The presence of both compressive and tensile regions, along with their evolution over time, demonstrates the influence of non-uniform temperature fields and the delayed response associated with fractional heat conduction.

\subsection{Analysis of Crack-Tip Response and Stress Intensity Factor}
\label{Analysis of Crack-Tip Response and Stress Intensity Factor}
\begin{figure}[htbp]
\centering
\begin{subfigure}[b]{0.47\textwidth}
\centering
\includegraphics[width=\textwidth]{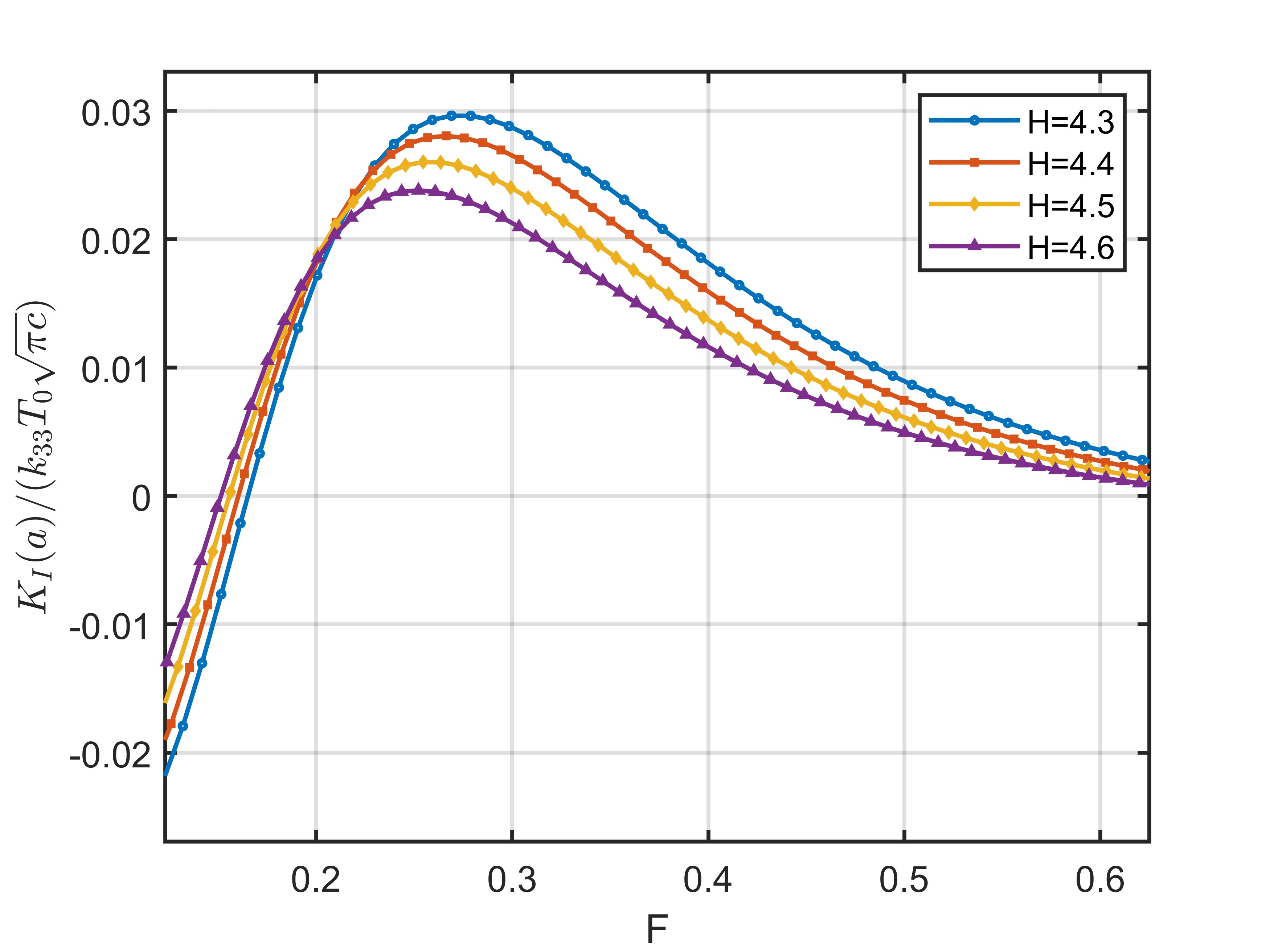}
\caption{Effect of strip thickness $H$}
\label{fig:5a}
\end{subfigure}
\hfill
\begin{subfigure}[b]{0.47\textwidth}
\centering
\includegraphics[width=\textwidth]{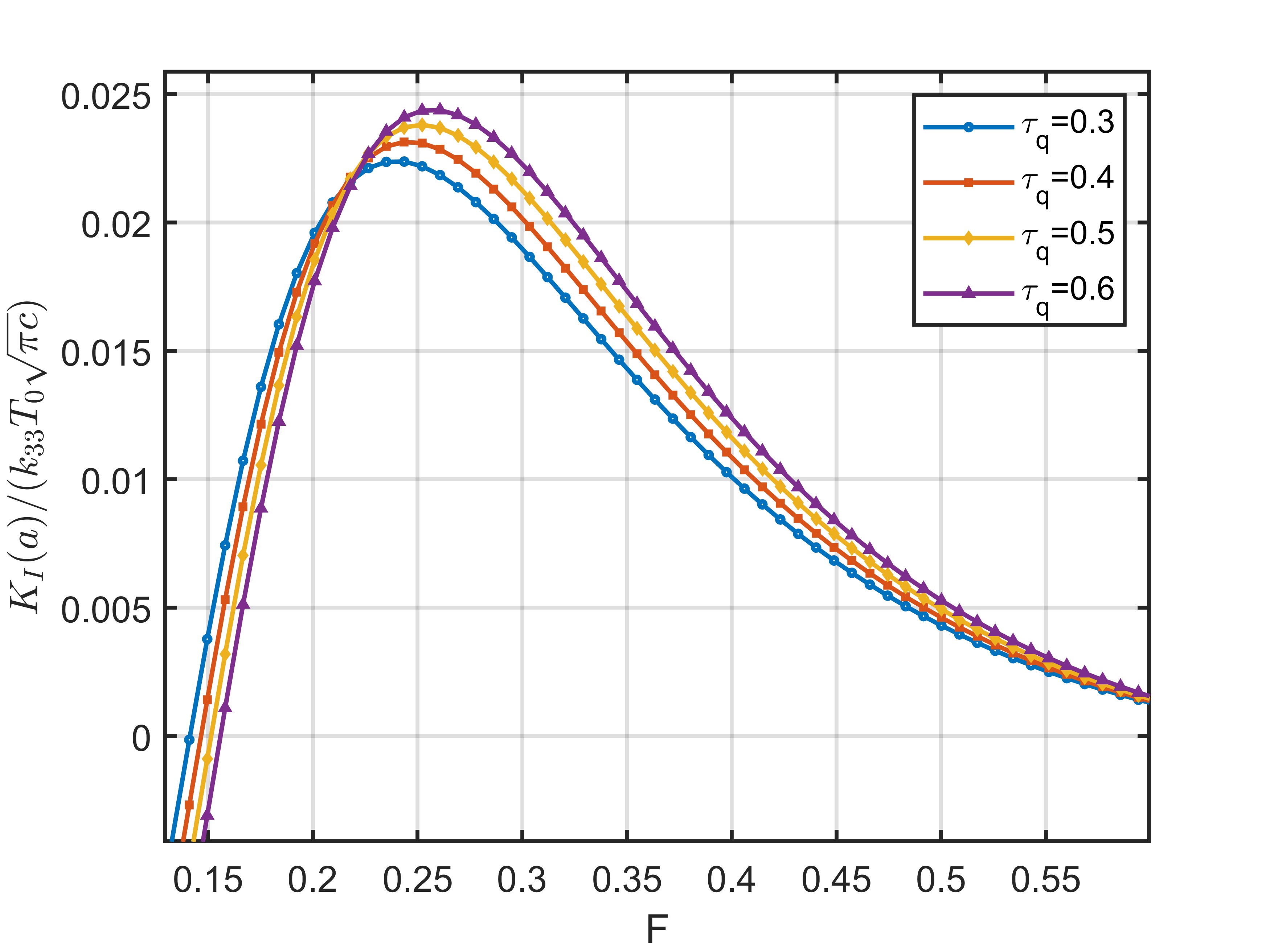}
\caption{Effect of relaxation time $\tau_q$}
\label{fig:5b}
\end{subfigure}
\vspace{0.4cm}
\begin{subfigure}[b]{0.47\textwidth}
\centering
\includegraphics[width=\textwidth]{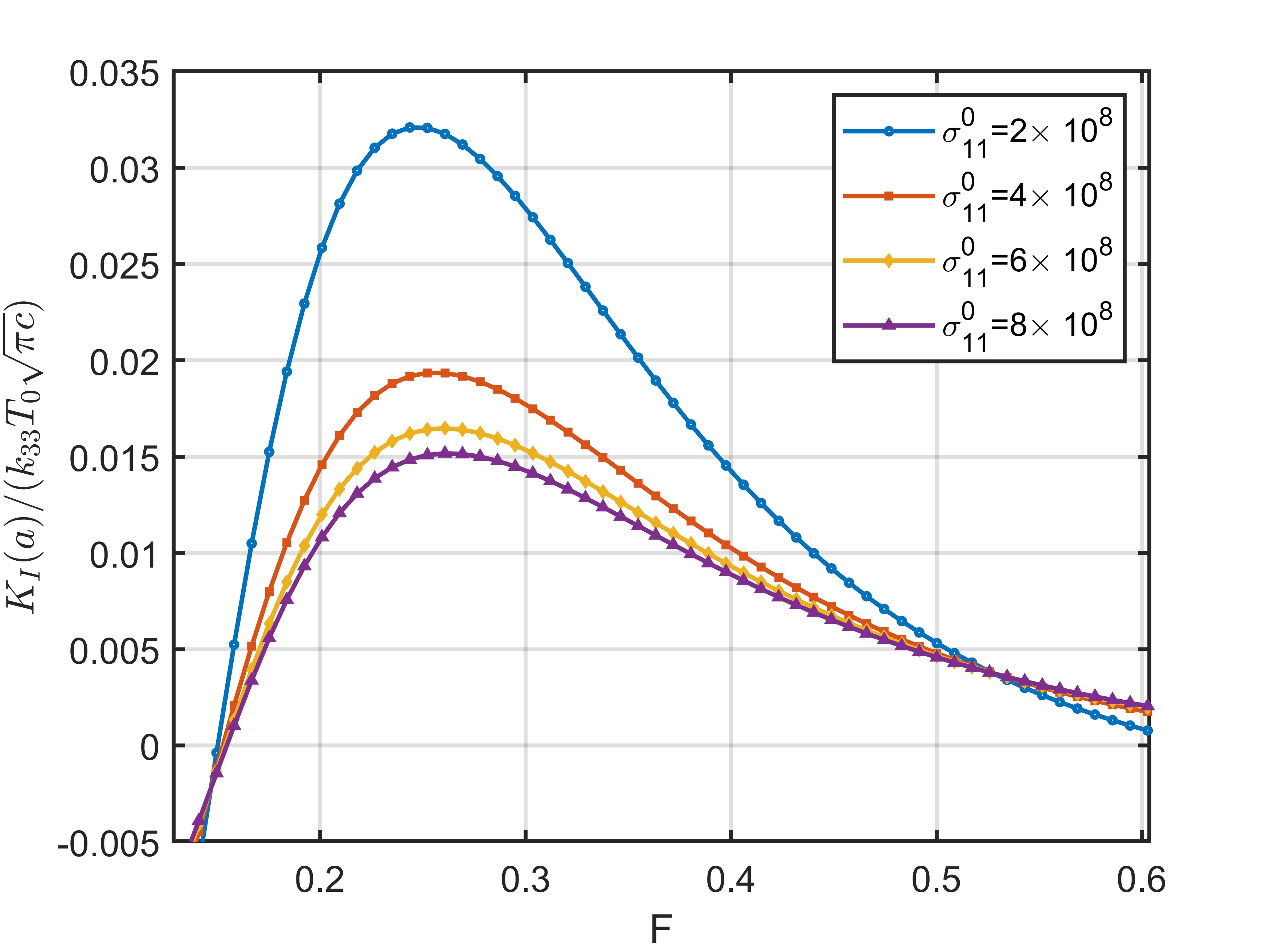}
\caption{Effect of initial stress $\sigma_{11}^0$}
\label{fig:5c}
\end{subfigure}
\hfill
\begin{subfigure}[b]{0.47\textwidth}
\centering
\includegraphics[width=\textwidth]{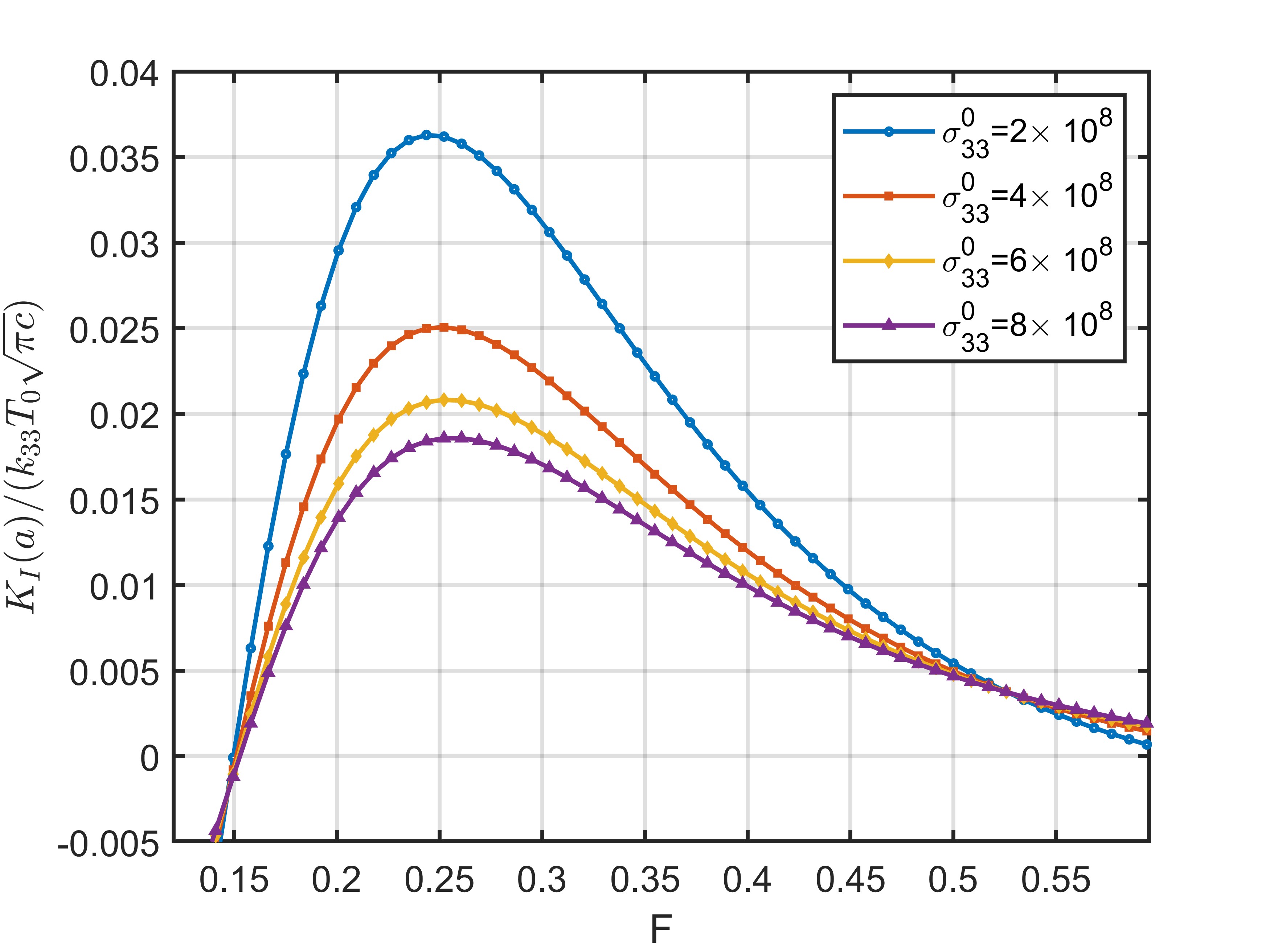}
\caption{Effect of initial stress $\sigma_{33}^0$}
\label{fig:5d}
\end{subfigure}

\caption{Variation of nondimensional stress intensity factor $K_I(a)/(k_{33}T_0\sqrt{\pi c})$ with respect to the Fourier number $F$ at $x_3 = a$ for different values of strip thickness $H$, thermal relaxation time $\tau_q$, and initial stresses $\sigma_{11}^0$ and $\sigma_{33}^0$.}
\label{fig:5}
\end{figure}

Figure~\ref{fig:5} shows the variation of the nondimensional stress intensity factor $K_I(a)/(k_{33}T_0\sqrt{\pi c})$ with respect to the Fourier number $F$ at $x_3=a$ for different values of strip thickness $H$, thermal relaxation time $\tau_q$, and initial stresses $\sigma_{11}^0$ and $\sigma_{33}^0$. It is observed that the stress intensity factor initially increases with $F$, attains a peak value, and subsequently decreases.
From Fig.~\ref{fig:5a}, it is observed that increasing the strip thickness $H$ reduces the peak value of the stress intensity factor, indicating a decrease in stress concentration near the lower crack tip. Figure~\ref{fig:5b} shows that an increase in the thermal relaxation time $\tau_q$ slightly enhances the peak value and shifts it towards higher values of $F$, reflecting the delayed thermal response associated with non-Fourier heat conduction.
The influence of initial stresses is presented in Figs.~\ref{fig:5c} and \ref{fig:5d}. It is observed that increasing $\sigma_{11}^0$ significantly decreases the magnitude of the stress intensity factor, indicating a reduction in crack driving force. A similar trend is observed for $\sigma_{33}^0$, although the variation is comparatively smoother, suggesting a relatively moderate influence on crack-tip behavior.

\begin{figure}[htbp]
\centering
\begin{subfigure}[b]{0.47\textwidth}
\centering
\includegraphics[width=\textwidth]{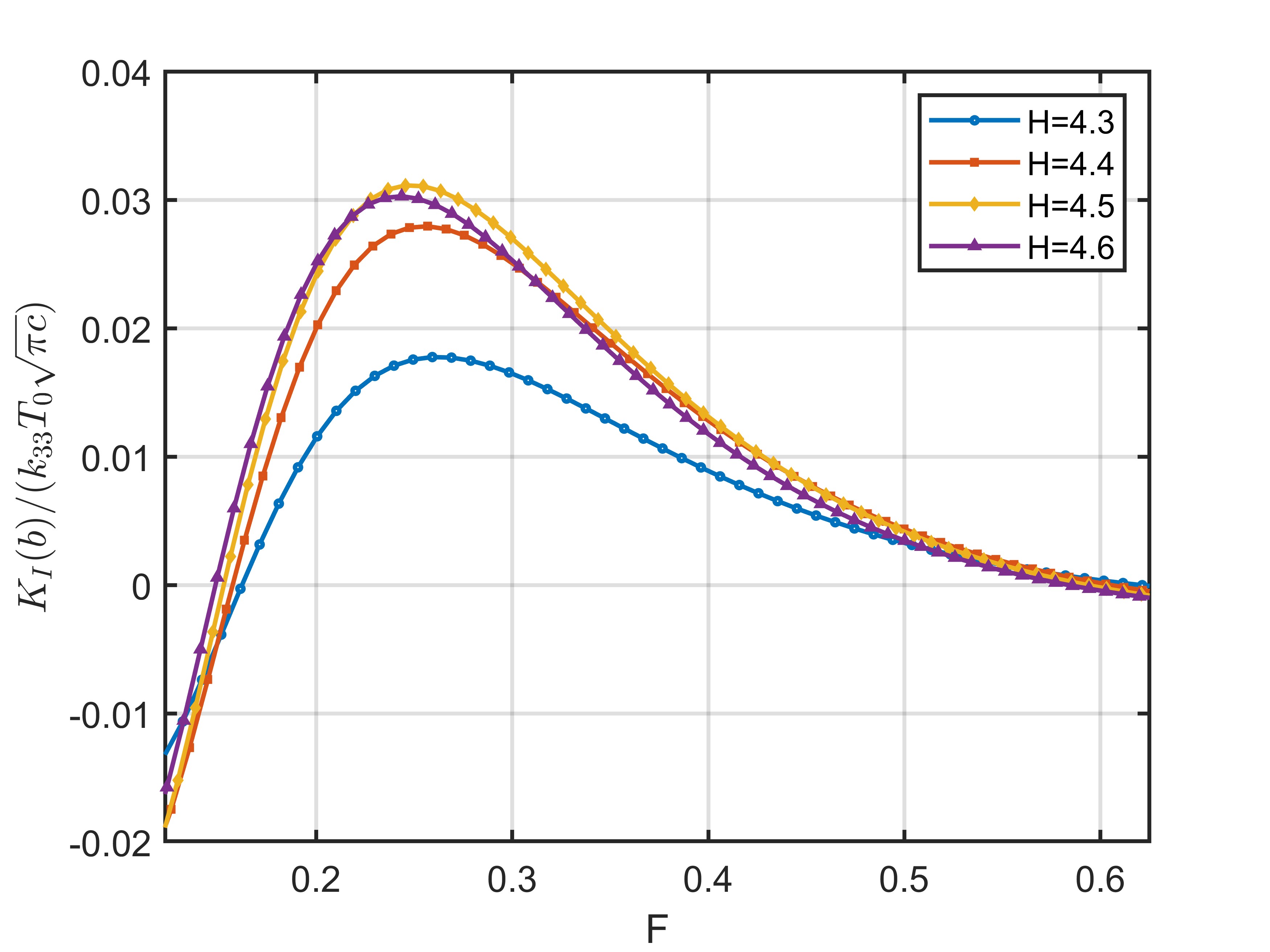}
\caption{Effect of strip thickness $H$}
\label{fig:6a}
\end{subfigure}
\hfill
\begin{subfigure}[b]{0.47\textwidth}
\centering
\includegraphics[width=\textwidth]{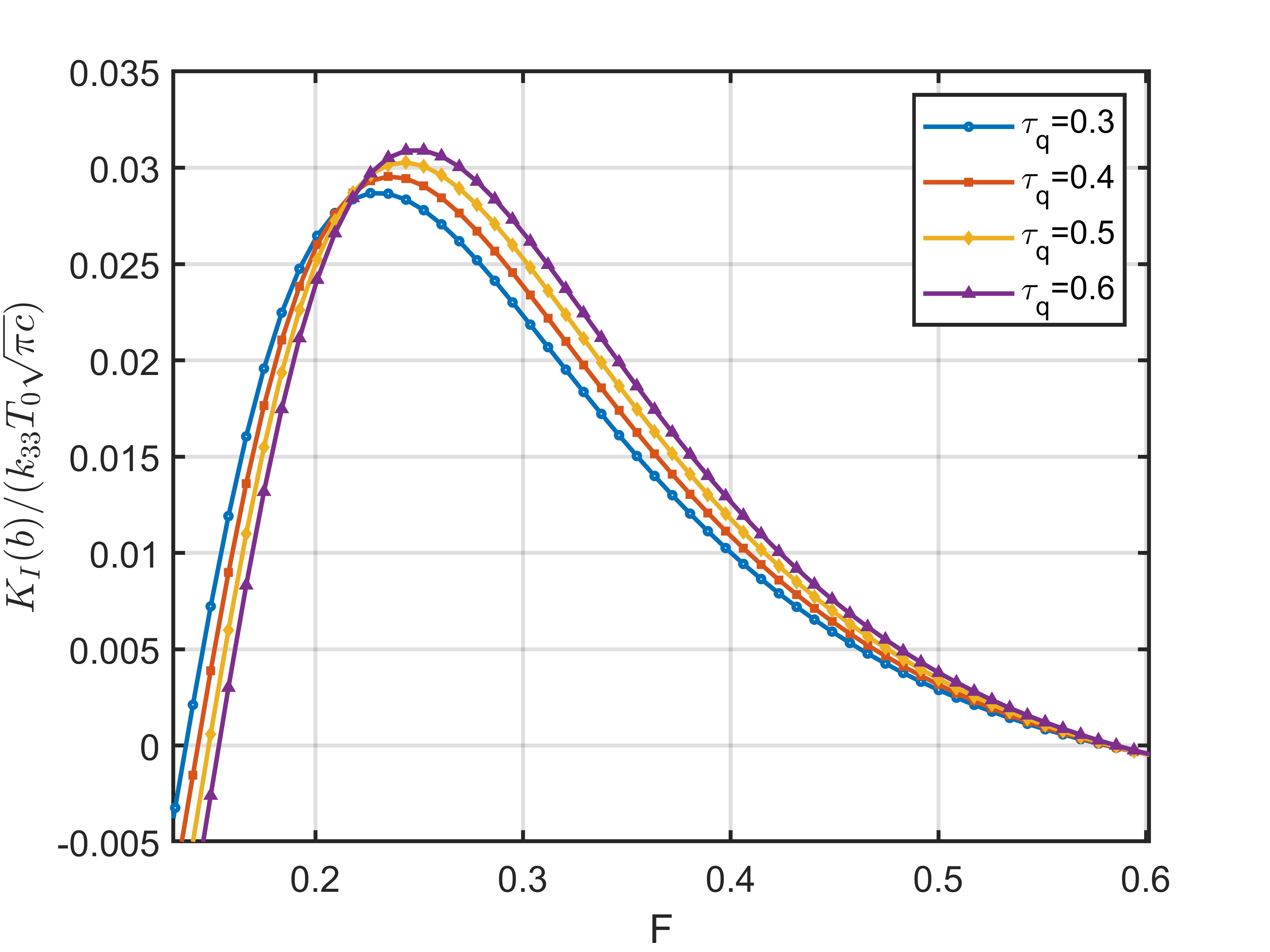}
\caption{Effect of relaxation time $\tau_q$}
\label{fig:6b}
\end{subfigure}
\vspace{0.4cm}
\begin{subfigure}[b]{0.47\textwidth}
\centering
\includegraphics[width=\textwidth]{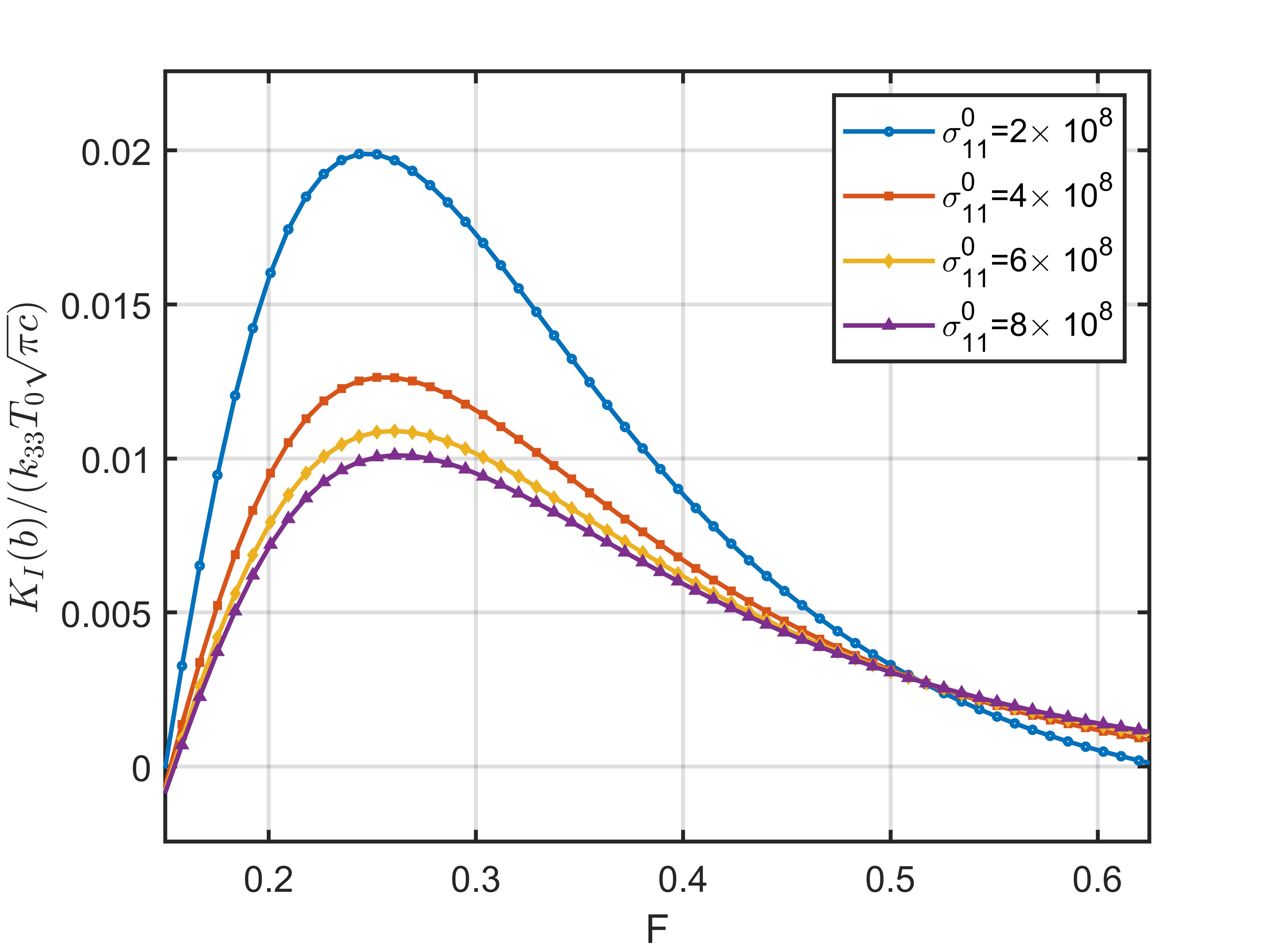}
\caption{Effect of initial stress $\sigma_{11}^0$}
\label{fig:6c}
\end{subfigure}
\hfill
\begin{subfigure}[b]{0.47\textwidth}
\centering
\includegraphics[width=\textwidth]{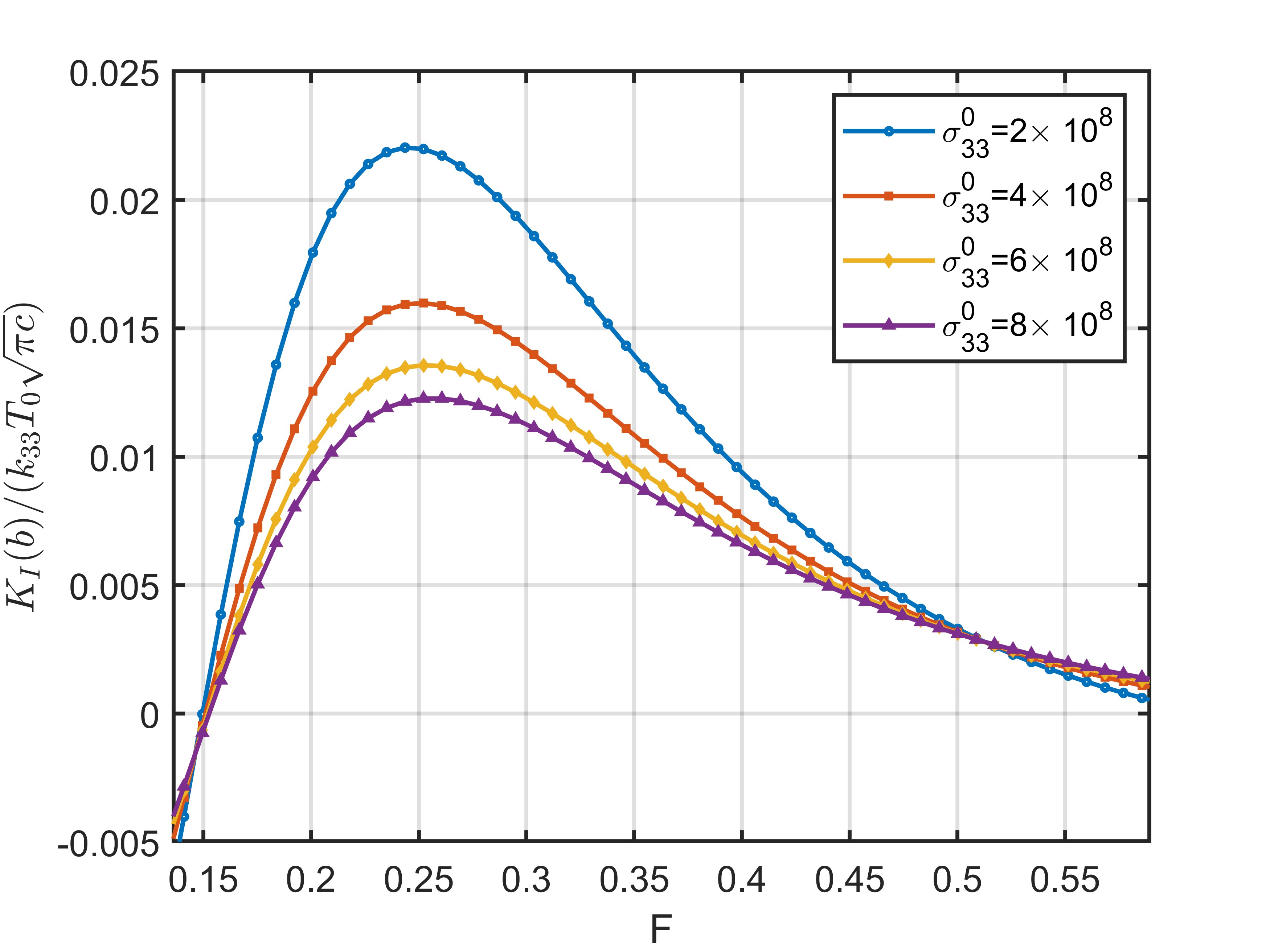}
\caption{Effect of initial stress $\sigma_{33}^0$}
\label{fig:6d}
\end{subfigure}

\caption{Variation of nondimensional stress intensity factor $K_I(b)/(k_{33}T_0\sqrt{\pi c})$ with respect to the Fourier number $F$ at $x_3 = b$ for different values of strip thickness $H$, thermal relaxation time $\tau_q$, and initial stresses $\sigma_{11}^0$ and $\sigma_{33}^0$.}
\label{fig:6}
\end{figure}

Figure~\ref{fig:6} presents the corresponding variation of the nondimensional stress intensity factor $K_I(b)/(k_{33}T_0\sqrt{\pi c})$ at the upper crack tip $x_3=b$. A similar trend is observed, in which the stress intensity factor increases with $F$, reaches a peak, and then gradually decreases.
However, the effect of strip thickness differs from that at the lower tip. Figure~\ref{fig:6a} shows that increasing $H$ leads to a higher peak stress intensity factor, indicating enhanced stress concentration at the upper crack tip. Figure~\ref{fig:6b} shows that increasing $\tau_q$ slightly increases the peak value and shifts it towards higher values of $F$, consistent with the delayed thermal response.
The influence of initial stresses is illustrated in Figs.~\ref{fig:6c} and \ref{fig:6d}. As $\sigma_{11}^0$ and $\sigma_{33}^0$ increase, the stress intensity factor decreases, indicating a stabilizing effect on crack propagation at the upper tip.

The stress intensity factor exhibits a pronounced peak with increasing Fourier number, indicating the existence of a critical transient regime at which the crack-tip response attains its maximum. Furthermore, the contrasting behavior observed between the lower and upper crack tips clearly demonstrates the non-uniform distribution of thermally induced stresses along the crack length.

\subsection{Special Case: Comparison with Classical Fourier Model}
\label{Special Case: Comparison with Classical Fourier Model}
\begin{figure}[htbp]
\centering

\begin{subfigure}[b]{0.47\textwidth}
\centering
\includegraphics[width=\textwidth]{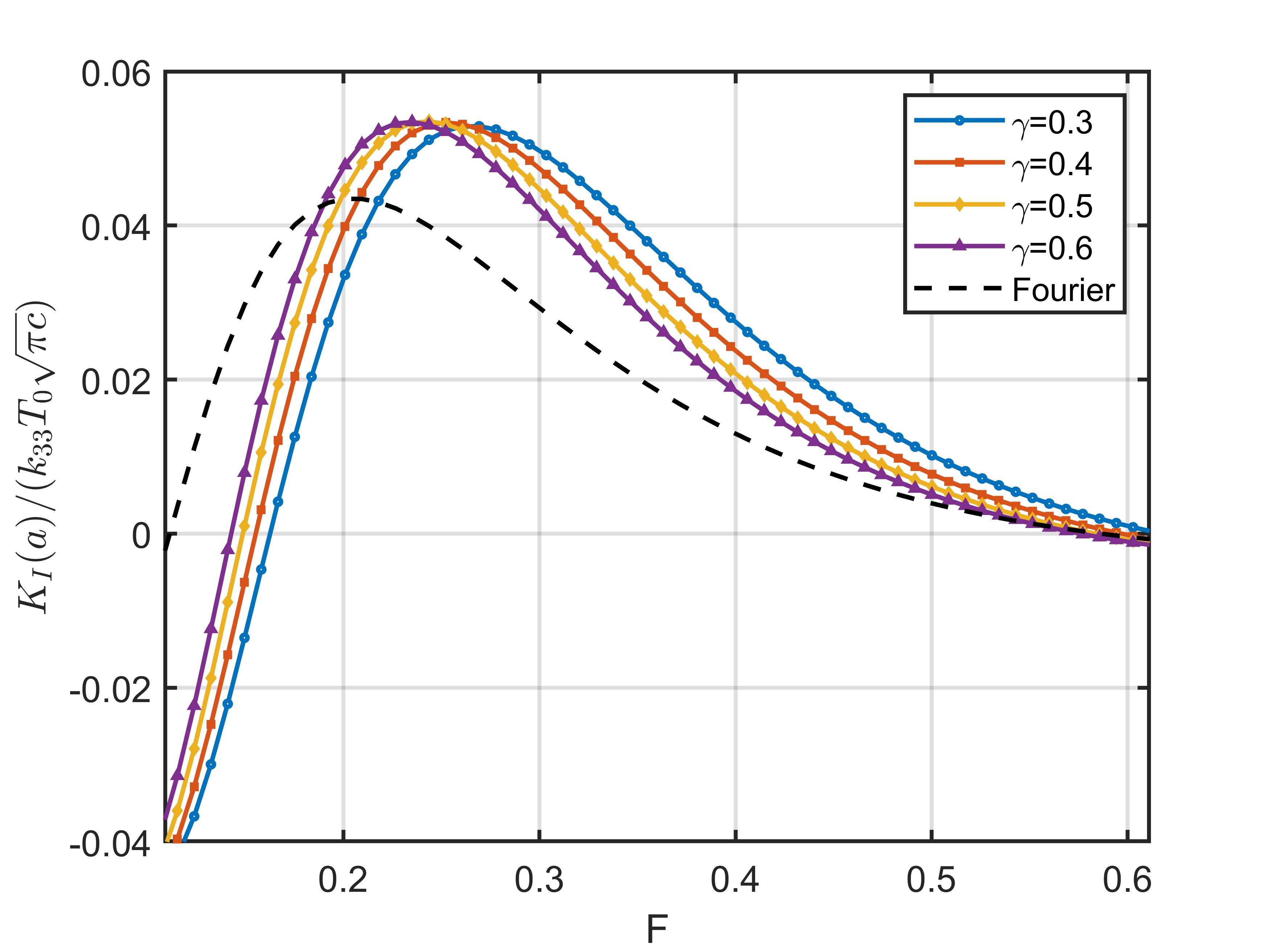}
\caption{Left crack tip}
\label{fig:7a}
\end{subfigure}
\hfill
\begin{subfigure}[b]{0.47\textwidth}
\centering
\includegraphics[width=\textwidth]{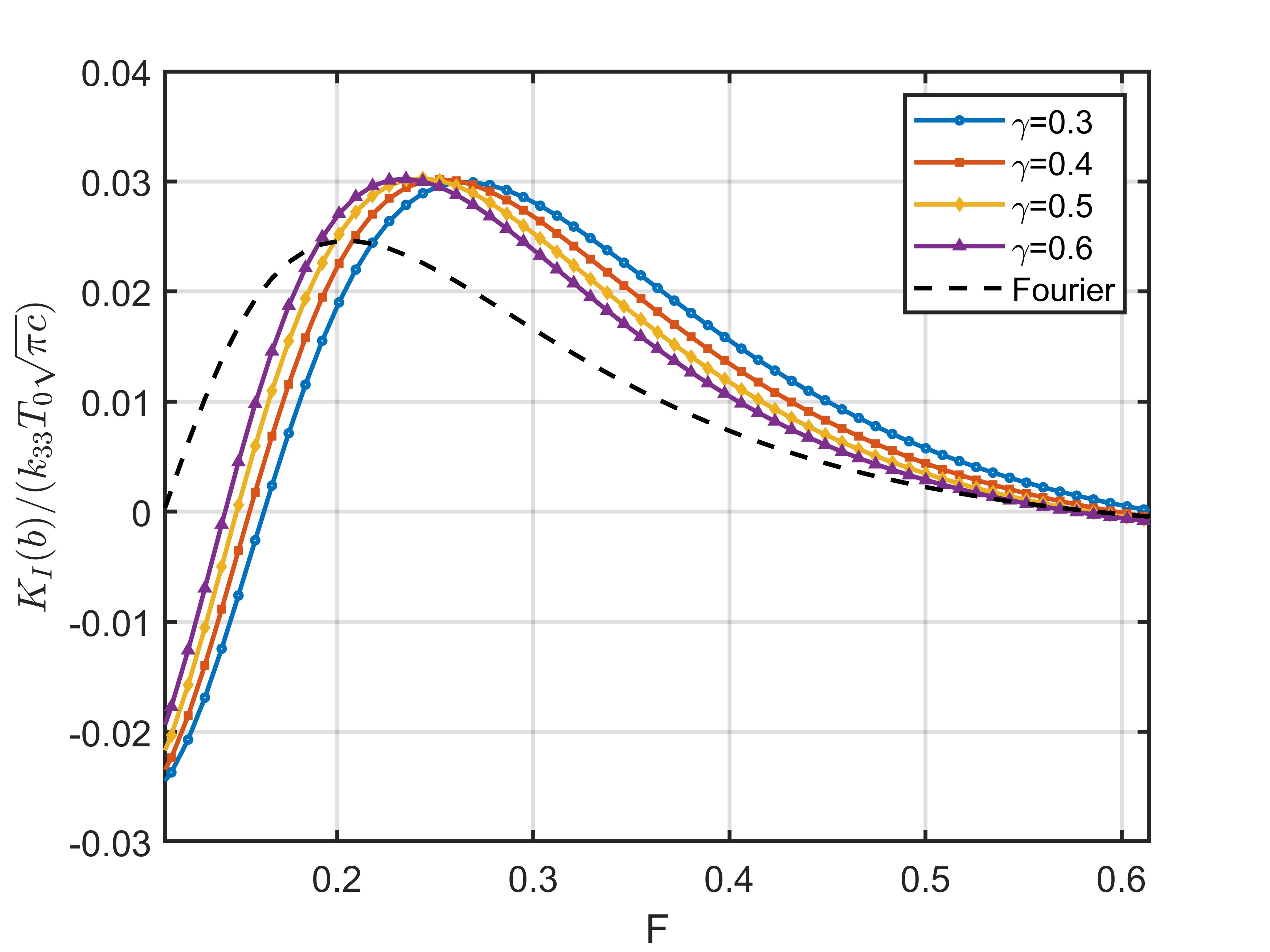}
\caption{Right crack tip}
\label{fig:7b}
\end{subfigure}

\caption{Variation of nondimensional stress intensity factor $K_I/(k_{33}T_0\sqrt{\pi c})$ with respect to the Fourier number $F$ at the crack tips for different values of fractional order $\gamma$.}
\label{fig:7}
\end{figure}

Figure~\ref{fig:7} presents the variation of the nondimensional stress intensity factor $K_I/(k_{33}T_0\sqrt{\pi c})$ with the Fourier number $F$ at both crack tips for various values of the fractional order $\gamma$. The present analysis is carried out for a fixed thermal relaxation time $\tau_q = 0.5$.
At the crack tip $x_3=a$(Fig.~\ref{fig:7a}), the stress intensity factor increases from negative values, attains a maximum within the range $F \approx 0.2$–$0.3$, and subsequently decreases towards zero as $F$ increases further. It is observed that increasing the fractional order from $\gamma = 0.3$ to $\gamma = 0.6$ leads to an increase in the peak value of the stress intensity factor, accompanied by a slight shift of the peak towards higher values of $F$. This behaviour indicates that increasing $\gamma$ weakens the memory effect and enhances the transient crack-tip response.
A similar trend is observed at the crack tip $x_3=b$ (Fig.~\ref{fig:7b}), although the magnitude of the stress intensity factor is comparatively smaller, reflecting the non-uniform distribution of thermal stresses along the crack length.

For comparison, the classical Fourier model is also included. It is observed that the Fourier solution predicts an earlier peak at lower values of $F$ with a reduced magnitude. This is because the classical Fourier law does not incorporate thermal relaxation time or fractional-order effects, and hence assumes instantaneous heat propagation without memory. In contrast, the present formulation incorporates both fractional heat conduction and a finite thermal relaxation time, which introduces non-local and time-delay effects in the heat transfer process.
Thus, the classical Fourier model can be regarded as a limiting case of the present formulation when the relaxation effects vanish and the heat conduction becomes purely local. The results clearly demonstrate that the combined influence of fractional order and thermal relaxation time significantly influences the transient crack-tip behavior.
\newpage
\section{Conclusion}
\label{Conclusion}
The propagation of thermoelastic fields and the associated stress intensity factors in a PZT-4 piezoelectric ceramic strip containing an internal vertical crack has been investigated within the framework of fractional heat conduction with thermal relaxation. The governing equations are formulated in the Laplace domain and solved numerically, while the corresponding time-domain responses are obtained using the Stehfest inversion technique. In addition, the resulting singular integral equation is solved using the Lobatto–Chebyshev method.
The influence of key material and physical parameters, including the fractional order, thermal relaxation time, Fourier number, strip thickness, and initial stresses, on the stress intensity factor and the associated thermoelastic fields is systematically analyzed. The principal findings of the present study are summarized as follows:
\begin{itemize}

\item The temperature distribution is strongly influenced by the fractional order and thermal relaxation time, 
and exhibits a pronounced decay along the strip thickness, indicating attenuation of thermal energy away from the heated boundary.

\item Higher fractional order accelerates thermal propagation, indicating reduced memory effects and enhanced energy transport, 
whereas increased thermal relaxation time introduces delay, reflecting non-Fourier heat conduction.

\item The temperature distribution is also affected by geometric and temporal parameters; increasing strip thickness smoothens the temperature field, 
while higher Fourier number leads to a more gradual spatial decay, indicating approach toward thermal equilibrium.

\item The thermal stress exhibits a transition from compressive to tensile behavior along the strip thickness, 
reflecting stress reversal due to non-uniform thermal expansion.

\item The thermal stress shows strong spatial and temporal variation, with pronounced concentration near the heated boundary at early times, 
followed by redistribution and gradual relaxation as the system evolves toward equilibrium.

\item The stress field is significantly influenced by material and relaxation parameters; increasing thermal relaxation time restricts 
the penetration of thermal disturbances, while the thermal modulus ratio alters both the magnitude and distribution of stress, 
highlighting the role of anisotropy.

\item The stress intensity factor near the crack tips varies nonlinearly with the Fourier number, reaching a maximum at a critical transient state 
and subsequently decreasing, with the peak shifting under thermal relaxation effects.

\item Increasing strip thickness reduces the stress intensity factor at the lower crack tip, while enhancing it at the upper crack tip, 
indicating asymmetric crack-tip response.

\item The initial stresses $\sigma_{11}^0$ and $\sigma_{33}^0$ reduce the magnitude of the stress intensity factor, 
indicating a stabilizing influence on crack propagation.

\item The stress intensity factor exhibits a pronounced peak with increasing Fourier number, indicating a critical transient regime, 
while the contrasting behavior at the two crack tips reflects a non-uniform distribution of thermally induced stresses along the crack length.

\item In comparison with the classical Fourier model, the fractional formulation predicts a higher and delayed peak in the stress intensity factor, 
reflecting memory-dependent and nonlocal heat conduction, while the Fourier model appears as a limiting case corresponding to instantaneous thermal propagation.

\item The results indicate that PZT-4 piezoelectric ceramics exhibit strong sensitivity to thermo-mechanical loading, 
supporting their applicability in smart structures and sensing devices.

\end{itemize}
The present formulation provides a consistent framework for analyzing transient thermoelastic fracture 
in piezoelectric ceramics and offers useful insights for the design of structures operating under 
coupled thermal and mechanical environments.

\noindent {\bf Acknowledgements}\\
The authors express their sincere gratitude to the National Institute of Technology Hamirpur for providing research facilities to Ms.~Diksha during her doctoral studies. The authors also acknowledge the University Grants Commission (UGC) for providing the research fellowship.\\
\noindent {\bf Conflicts of interest}\\
The authors declare no conflict of interest.\\

\appendix
\section{}
\label{sec:eigenvectors}
Substitution of the general solutions into the transformed governing equations yields the following homogeneous algebraic systems for the eigenvectors associated with the characteristic roots $\Upsilon_{2j}$ and $\Upsilon_{1j}$.
\begin{equation}
\begin{bmatrix}
\mu_{11}+\sigma_{11}^0-(\mu_{44}+\sigma_{33}^0)\Upsilon_{2j}^{2} &
(\mu_{13}+\mu_{44})\Upsilon_{2j} &
(e_{31}+e_{15})\Upsilon_{2j}
\\[6pt]

(\mu_{13}+\mu_{44})\Upsilon_{2j} &
(\mu_{33}+\sigma_{33}^0)\Upsilon_{2j}^{2}-(\mu_{44}+\sigma_{11}^0) &
e_{33}\Upsilon_{2j}^{2}-e_{15}
\\[6pt]

(e_{31}+e_{15})\Upsilon_{2j} &
e_{33}\Upsilon_{2j}^{2}-e_{15} &
\varepsilon_{11}-\varepsilon_{33}\Upsilon_{2j}^{2}
\end{bmatrix}
\begin{bmatrix}
M_{1j}\\
M_{2j}\\
M_{3j}
\end{bmatrix}
=
\begin{bmatrix}
0\\
0\\
0
\end{bmatrix}.
\end{equation}
\begin{equation}
\begin{bmatrix}
(\mu_{44}+\sigma_{33}^0)-(\mu_{11}+\sigma_{11}^0)\Upsilon_{1j}^{2} &
i\dfrac{|p|}{p}(\mu_{13}+\mu_{44})\Upsilon_{1j} &
i\dfrac{|p|}{p}(e_{31}+e_{15})\Upsilon_{1j}
\\[8pt]

i\dfrac{|p|}{p}(\mu_{13}+\mu_{44})\Upsilon_{1j} &
(\mu_{33}+\sigma_{33}^0)-(\mu_{44}+\sigma_{11}^0)\Upsilon_{1j}^{2} &
e_{33}-e_{15}\Upsilon_{1j}^{2}
\\[8pt]

i\dfrac{|p|}{p}(e_{31}+e_{15})\Upsilon_{1j} &
e_{33}-e_{15}\Upsilon_{1j}^{2} &
-\varepsilon_{33}+\varepsilon_{11}\Upsilon_{1j}^{2}
\end{bmatrix}
\begin{bmatrix}
N_{1j}\\
N_{2j}\\
N_{3j}
\end{bmatrix}
=
\begin{bmatrix}
0\\
0\\
0
\end{bmatrix}.
\end{equation}
For nontrivial solutions, the determinants of the above coefficient matrices must vanish, which yields the characteristic equations governing the roots $\Upsilon_{2j}$ and $\Upsilon_{1j}$.
As the stresses and electric displacements vanish as $x_{1} \to \infty$, 
the admissible roots $\Upsilon_{1j}$ $(j = 1,2,3)$ are chosen such that
\[
\operatorname{Re}(\Upsilon_{1j}) < 0 ,
\]
ensuring that the corresponding exponential terms decay away from the crack region.

\section{}
\label{dikj}
The coefficients $d_{ikj}(p)$ $(i=1,2,\ldots,5)$ appearing in the stress and electric–displacement expressions are defined as follows.
\begin{align}
d_{12j} &= \mu_{11}M_{1j}
          + \mu_{13}\Upsilon_{2j} M_{2j}
          + e_{31}\Upsilon_{2j} M_{3j}, \\[4pt]
d_{22j} &= e_{15}\Upsilon_{2j} M_{1j}
          - e_{15}M_{2j}
          + \varepsilon_{11}M_{3j}, \\[4pt]
d_{32j} &= \mu_{44}\Upsilon_{2j} M_{1j}
          - \mu_{44}M_{2j}
          - e_{15}M_{3j}, \\[4pt]
d_{42j} &= \mu_{13}M_{1j}
          + \mu_{33}\Upsilon_{2j} M_{2j}
          + e_{33}\Upsilon_{2j} M_{3j}, \\[4pt]
d_{52j} &= e_{31}M_{1j}
          + e_{33}\Upsilon_{2j} M_{2j}
          - \varepsilon_{33}\Upsilon_{2j} M_{3j}.
\end{align}
\begin{align}
d_{11j} &= \mu_{11}\frac{|p|}{p}\Upsilon_{1j} N_{1j}
          - i\,\mu_{13}N_{2j}
          - i\,e_{31}N_{3j}, \\[4pt]
d_{21j} &=  -i\,e_{15}N_{1j}
          + e_{15}\frac{|p|}{p}\Upsilon_{1j} N_{2j}
          - \varepsilon_{11}\frac{|p|}{p}\Upsilon_{1j} N_{3j},\\[4pt]
d_{31j} &= -i\,\mu_{44}N_{1j}
          + \mu_{44}\frac{|p|}{p}\Upsilon_{1j} N_{2j}
          + e_{15}\frac{|p|}{p}\Upsilon_{1j} N_{3j}, \\[4pt]
d_{41j} &= \mu_{13}\frac{|p|}{p}\Upsilon_{1j} N_{1j}
          - i\,\mu_{33}N_{2j}
          - i\,e_{33}N_{3j}, \\[4pt]
d_{51j} &= e_{31}\frac{|p|}{p}\Upsilon_{1j} N_{1j}
          - i\,e_{33}N_{2j}
          + i\,\varepsilon_{33}N_{3j}.
\end{align}

\section{}
\label{matrix theta}
The coefficients $\psi_j(p,r)$ $(j=1,2,\ldots,6)$ are obtained from the following system of linear algebraic equations:
\begin{equation}
\begin{bmatrix}
\psi_1 \\
\psi_2 \\
\psi_3 \\
\psi_4 \\
\psi_5 \\
\psi_6
\end{bmatrix}
=
\mathbf{\Theta}^{-1}
\begin{bmatrix}
\Phi_1 \\
\Phi_2 \\
\Phi_3 \\
\Phi_4 \\
\Phi_5 \\
\Phi_6
\end{bmatrix},
\end{equation}
where the coefficient matrix $\mathbf{\Theta}$ is given by
\begin{equation}
\mathbf{\Theta}=
\begin{bmatrix}
d_{321}e^{p\Upsilon_{21} H} &
d_{322}e^{p\Upsilon_{22} H} &
d_{323}e^{p\Upsilon_{23} H} &
d_{324}e^{p\Upsilon_{24} H} &
d_{325}e^{p\Upsilon_{25} H} &
d_{326}e^{p\Upsilon_{26} H}
\\[6pt]
d_{321} & d_{322} & d_{323} & d_{324} & d_{325} & d_{326}
\\[6pt]

d_{421}e^{p\Upsilon_{21} H} &
d_{422}e^{p\Upsilon_{22} H} &
d_{423}e^{p\Upsilon_{23} H} &
d_{424}e^{p\Upsilon_{24} H} &
d_{425}e^{p\Upsilon_{25} H} &
d_{426}e^{p\Upsilon_{26} H}
\\[6pt]
d_{421} & d_{422} & d_{423} & d_{424} & d_{425} & d_{426}
\\[6pt]

d_{521}e^{p\Upsilon_{21} H} &
d_{522}e^{p\Upsilon_{22} H} &
d_{523}e^{p\Upsilon_{23} H} &
d_{524}e^{p\Upsilon_{24} H} &
d_{525}e^{p\Upsilon_{25} H} &
d_{526}e^{p\Upsilon_{26} H}
\\[6pt]

d_{521} & d_{522} & d_{523} & d_{524} & d_{525} & d_{526}
\end{bmatrix}.
\label{eq:Theta_matrix}
\end{equation}
The functions $\Phi_i(p,r)$ $(i=1,2,\ldots,6)$ are obtained as
\begin{align}
\Phi_1(p,r)&= -\sum_{j=1}^{3}\int_{-\infty}^{\infty}\frac{p}{\xi^{2}\Upsilon_{1j}^{2}+p^{2}}d_{31j} \lambda_j\,e^{i\xi(r-H)}\, d\xi
\\[6pt]
\Phi_2(p,r)&= -\sum_{j=1}^{3}\int_{-\infty}^{\infty}\frac{p}{\xi^{2}\Upsilon_{1j}^{2}+p^{2}}d_{31j} \lambda_j\,e^{i\xi r}\, d\xi\\[6pt]
\Phi_3(p,r)&= \sum_{j=1}^{3}\int_{-\infty}^{\infty}
\frac{|\xi|\Upsilon_{1j}}{\xi^{2}\Upsilon_{1j}^{2}+p^{2}}d_{41j} \lambda_j\,e^{i\xi(r-H)}\, d\xi 
\\[6pt]
\Phi_4(p,r)&= \sum_{j=1}^{3}\int_{-\infty}^{\infty}\frac{|\xi|\Upsilon_{1j}}{\xi^{2}\Upsilon_{1j}^{2}+p^{2}}d_{41j} b_j\,
e^{i\xi r}\, d\xi
\\[6pt]
\Phi_5(p,r)&= \sum_{j=1}^{3}\int_{-\infty}^{\infty}\frac{|\xi|\Upsilon_{1j}}{\xi^{2}\Upsilon_{1j}^{2}+p^{2}}d_{51j} \lambda_j\,
e^{i\xi(r-H)}\, d\xi 
\\[6pt]
\Phi_6(p,r)&= \sum_{j=1}^{3}\int_{-\infty}^{\infty}\frac{|\xi|\Upsilon_{1j}}{\xi^{2}\Upsilon_{1j}^{2}+p^{2}}d_{51j} \lambda_j\,
e^{i\xi r}\, d\xi 
\end{align}

\bibliographystyle{unsrt}
\bibliography{refrences}
\end{document}